\documentstyle[axodraw,12pt]{article}

\begin{document}

\hoffset = -1truecm
\voffset = -2truecm
\baselineskip = 10 mm

\title{\bf A geometric higher twist effect on nuclear target}

\author{
{\bf
Wei Zhu
}\\
\normalsize Department of Physics, East China Normal University,
\normalsize Shanghai 200062, {\bf P.R. China}
}

\date{}

\newpage

\maketitle

\vskip 3truecm

\begin{abstract}

        The higher twist effects in deep inelastic scattering
on the nuclear target are studied using time ordered perturbation
theory. We showed that the collinear rescattering of the outgoing
quark on the extra nucleons via the contacting gluon-pair is
dominant in nuclear size-dependent effects. The Qiu-Vitev
resummation is proved by using the geometric properties of the
higher twist amplitudes. The leading contributions of
nuclear-enhanced effect to the DGLAP evolution equation are
resummed in the same framework.

\end{abstract}


$keywords$: Higher twist effect; PQCD evolution equation

\newpage
\begin{center}
\section{Introduction}
\end{center}

    Deep inelastic scattering (DIS) experiments are the main
source extracting the information about QCD structure of the
hadronic target. The DIS amplitudes can be expanded according to
their twists. The leading-twist amplitudes describe the
quark-gluon structure of a target at short distance, while the
higher twist effects and their summation may provide a valuable
information for the multi-parton correlations at long distance in
the target [1]. The nuclear target is an ideal ``laboratory" for
testing higher twist theory because of its special multi-parton
configuration.

     However, it is difficult to sum all contributions in the twist
expansion due to the complicated structure of the higher twist
amplitudes and the unknown multi-parton correlation functions.
Instead of a complete summation, the resummation of the leading
contributions to a special effect from the twist expansion is
useful. For example, the Glauber model is used to resum the
contributions of the multi-parton scattering to the shadowing
effect in the nuclear structure functions in the Glauber-Mueller
formula [2]. This formula assumes that the leading contributions
are from the incoherent scattering of color dipole on the nuclear
gluons.  Another example in this realm is the
Jalilian-Marian--Iancu--McLerran--Weigert--Leonidov--Kovner
(JIMWLK) equation [3], where the soft nuclear gluons are treated
as a classical color potential because of its large density at
small $x_B$. In this background, energetic partons in the
multi-dipole configuration interact with soft gluons through the
Wilson lines, which resum the contributions of the soft nuclear
gluons at the eikonal approximation [4,5]. Both of the above two
researches are broadly used to predict the saturation phenomena in
the nuclear structure functions, where unitarity is restored.

    On the other hand, a different resummation of the nuclear
higher twist amplitudes is recently proposed by Qiu and Vitev
(Qiu-Vitev resummation) [6]. This model assumes that the leading
contributions to the nuclear size effect in the structure
functions originate from the singularities of the gluon
correlation functions in the higher twist amplitudes [7].

    The higher twist amplitudes have complicated color-spin
structure. The resummation of a higher twist effect is valid only
if the contributions from a lot of irrelevant diagrams can be
$entirely$ neglected. Unfortunately, this point is unclear in the
researches mentioned above.

    The motivation of the present work is to study the geometric
properties of the higher twist amplitudes for the DIS on nuclear
target using time ordered perturbation theory (TOPT) [8], which
was developed in the collinear approximation (CLTOPT) in [9].
Comparing with the standard covariant perturbation theory (CVPT),
the CLTOPT provides a more intuitive geometric character of the
higher twist amplitudes. Using the above TOPT method we study the
DIS processes on the nuclear target at arbitrary twist. We present
a geometric higher twist effect in the DIS on the nuclear target:
the leading contributions to a nuclear size-dependent effect in
the DIS processes originate from the multiple rescattering of the
outgoing quark on the extra nucleon via the exchange of the
contacting gluon-pair. It reproduces the Qiu-Vitev resummation [6]
but using different dynamics. In particular, this space-time
character of the amplitudes in the CLTOPT allows us to distinguish
the leading diagrams from a lot of sub-leading diagrams without
calculations of the Feynman diagrams. We resum the corrections of
the geometric higher twist effect to the structure function and
DGLAP evolution equation [10,11].

         The paper is organized as follows. In Section 2 we present
a short review about TOPT in the collinear approximation, which is
necessary for understanding this work. For a complete calculation
of the higher twist amplitude including the matrix element, we
discuss in Section 3 the normalization of the correlation
functions. In Section 4, using CLTOPT we demonstrate the
space-time picture of the higher twist amplitudes, which exposes
obvious nuclear size-dependence. In Section 5 we derive the
expressions for the nuclear dependence of the structure function
at twist-4 level, then the results are generalized to sum the
contributions of all twist amplitudes. In Section 6 we study the
corrections of the above mentioned higher twist effect to the
DGLAP evolution equation. The discussions and a summary are
presented in Section 7.

\newpage
\begin{center}
\section{Definitions and tools}
\end{center}

        We consider DIS of leptons from an unpolarized nuclear
target.  We denote the light-like vectors along the positive and
negative directions as $n^{\mu}\equiv 1/\sqrt{2}(1,0_{\perp},-1)$
and $\overline{n}^{\mu}\equiv 1/\sqrt{2}(1,0_{\perp},1).$
Therefore, we have

\begin{eqnarray}
n\cdot n=\overline{n}\cdot \overline{n}=0,~~\overline{n}\cdot
n=1.
\end{eqnarray}

     As is well known, the infinite momentum frame is equivalent
to the light-cone coordinate. Here, however, we would rather use
TOPT in the infinite momentum frame than the light-cone
quantization. The reason is that the contributions of the backward
(instantaneous) propagator in the latter case are neglected and it
is not a complete description for the higher twist processes. In
the concrete, we choose an infinite frame, in which the target
momentum and virtual photon momentum have the following forms,

\begin{eqnarray}
P^{\mu}&=& P^+\overline{n}^{\mu},\nonumber\\
q^{\mu}&=&
q^+\overline{n}+q^-n=-x_BP^+\overline{n}^{\mu}+\frac{Q^2}{2x_BP^+}n^{\mu},
\end{eqnarray}
where $Q^2=-q^2$ and $x_B$ is the Bjorken variable.

    In this work we take the physical axial gauge: $n\cdot A=0$, with $A$ being
the gluon field. The inclusive cross-section for the unpolarized
DIS is determined by the hadronic tensor $W^{\mu\nu}$, which can
be decomposed into the transverse structure function
$F_T(x_B,Q^2)$ and longitudinal one $F_L(x_B,Q^2)$,

\begin{equation}
W^{\mu\nu}(x_B,Q^2)=e_L^{\mu\nu}F_L(x_B,Q^2)+e_T^{\mu\nu}F_T(x_B,Q^2).
\end{equation}
The projection operators are defined as

\begin{eqnarray}
e_L^{\mu\nu} & = &
\frac{1}{2Q^2}\left(x_BP^+\overline{n}^{\mu}+\frac{Q^2}
{2x_BP^+}n^{\mu}\right) \left(x_BP^+\overline{n}^{\nu}+\frac{Q^2}
{2x_BP^+}n^{\nu}\right), \\
e_T^{\mu\nu} & = & \frac{1}{2}(\overline{n}^{\mu}n^{\nu}+
\overline{n}^{\nu}n^{\mu} -g^{\mu\nu})
\equiv \frac{1}{2}d^{\mu\nu}.
\end{eqnarray}
Clearly, using

\begin{equation}
e_T^{\mu\nu}\cdot e_T^{\mu\nu}=\frac{1}{2},~~e_L^{\mu\nu}
\cdot e_L^{\mu\nu}=\frac{1}{4}~~{\hbox{\rm and}}~~
e_T^{\mu\nu}\cdot e_L^{\mu\nu}=0,
\end{equation}
one can directly extract $F_T(x_B,Q^2)$ and $F_L(x_B,Q^2)$ from
$W^{\mu\nu}$ according to

\begin{eqnarray}
F_T(x_B,Q^2)&=&2e_T^{\mu\nu}W^{\mu\nu}(x_B,Q^2), \\
F_L(x_B,Q^2)&=&4e_L^{\mu\nu}W^{\mu\nu}(x_B,Q^2).
\end{eqnarray}

        Note that the hadronic tensor $W^{\mu\nu}(x_B,Q^2)$ can also be
decomposed into the structure functions $F_1(x_B,Q^2)$ and $F_2(x_B,Q^2)$
and their relations with Eqs. (7) and (8) are
\begin{eqnarray}
F_L(x_B,Q^2) & = & \frac{F_2(x_B,Q^2)}{x_B}-2F_1(x_B,Q^2),\nonumber\\
F_T(x_B,Q^2) & = & 2F_1(x_B,Q^2).
\end{eqnarray}

         To further examine the role of higher twist
corrections in the DIS processes, we present the TOPT-form of the
QCD-propagators in the collinear limit [9]. As will be discussed
in details below, the nuclear dependence on both the structure
functions and evolution equations is dominated by the higher twist
amplitudes. A straightforward reason is that the amplitude at
higher twist contains the multi-parton operators in the matrix
element, which can come from the different nucleons of a nucleus
and give rise to the nuclear size-dependence. However, we have a
lot of amplitudes corresponding to the multiple interactions among
the partons. The important question is: which of them contributes
the leading nuclear effect to the processes at a given order of
approximation? We find that the TOPT in the collinear limit, which
is developed in our previous works [9], is a useful tool for
answering the above question.

        TOPT is exactly equivalent to the method of the
Feynman diagram [1]. A covariant propagator in the Feynman diagram
is decomposed to the forward- and backward-propagators in TOPT and
this leads to two remarkable properties: the interaction is
ordered on time and the propagating momentum is on the mass shell.
As we will point out in the following, the former shows how a
process evolves in the light-cone coordinates and gives us
information about the nuclear effects, while the latter is useful
to decompose a complicated calculation.

    At first sight it seems like that TOPT leads to a
proliferation of diagrams. However, we find
that there is only one TOPT diagram corresponding to one Feynman diagram
in the collinear approximation, where the transverse momentum of the parton is
neglected. In fact, the TOPT-propagators in the collinear limit have
the following special forms [9]
\begin{eqnarray}
S_F(k)&=&\frac{\gamma\cdot \overline {n}}{2k\cdot\overline{n}}, \\
S_B(k)&=&\frac{\gamma\cdot n}{2k\cdot n},
\end{eqnarray}
for quarks, and

\begin{eqnarray}
G_F^{\alpha\beta}(k)&=&\frac{d^{\alpha\beta}_{\perp}}
{k\cdot \overline{n}(k\cdot n-k\cdot \overline{n})}\\
G_B^{\alpha\beta}(k)&=&\frac{n^{\alpha}n^{\beta}}{(k\cdot n)^2},
\end{eqnarray}
for gluons, where $``F"$ and $``B"$ label the forward- and
backward-moving partons, respectively.  Using helicity
conservation and Eq. (1), the non-vanishing vertices of QCD and
QED in the collinear approximation are shown in Figs. 1 and 2,
respectively. According to those selection rules, we draw the
twist-4 amplitudes, which contribute to $F_T(x_B,Q^2)$ and
$F_L(x_B,Q^2)$ in the collinear limit, in Figs. 3 and 4, where we
label the outgoing quark or outgoing gluon as $``B"$ or $``F"$,
since it contributes $\sim \gamma\cdot n$ or $\sim
d^{\alpha\beta}_{\perp}$, respectively. We did not draw the
conjugate diagrams. One finds that only one TOPT graph corresponds
to one covariant Feynman diagram in the collinear limit.

        The deep inelastic interaction in the TOPT-form is time-ordered in
a given infinite momentum frame and shows how a system evolves in
the light-cone coordinates. For illustrating this picture, we take
the Fourier transformation of the forward- and
backward-propagators. Note that the right hand sides of Eqs. (10)
and (11) indicate that $S_F$ is only a function of $k^-$, while
$S_B$ is related to $k^+$. Thus, we have

\begin{eqnarray}
\int S_F(k^-)\exp[ik\cdot (y_1-y_2)]\frac{d^4k}{(2\pi)^4}
 & = & f_F(y_1^+-y_2^+)\delta(y_1^--y_2^-)\delta^{(2)}(y_1^{\perp}-
y_2^{\perp})\nonumber\\
 & \equiv & S_F(y_1^+-y_2^+), \\
\int S_B(k^+)\exp[ik\cdot (y_1-y_2)]\frac{d^4k}{(2\pi)^4}
 & = & f_B(y_1^--y_2^-)\delta(y_1^+-y_2^+)\delta^{(2)}(y_1^\perp-y_2^\perp)
\nonumber\\
& \equiv & S_B(y_1^--y_2^-).
\end{eqnarray}
These expressions show that the forward- and backward-propagators
for quarks in the collinear approximation are parallel to $y^+$-
and $y^-$-axes, respectively. Using this property, for example, we
draw the twist-4 amplitudes of Fig. 3a in the $y^+-y^-$-coordinate
in Fig. 5a. One finds that $S_F$ and $S_B$ correspond to the
contact- and instantaneous-interactions in the light-cone
coordinates. The backward propagator of the gluon (Eq. (13)) has
the same property as $S_B$, i.e., $G_B(y_1^--y_2^-)$ is always
parallel to the $y^-$-axis. We focus on the relation between the
interactions of $\gamma^*$-parton and the structure of the
correlation functions, which are defined on the $y^-$-axis in the
collinear factorization. Therefore, for simplicity, we project the
interactions of $\gamma^*$-parton in Fig. 5a on the $y^-$-axis
(Fig. 5b). Thus, we get the geometric diagrams Figs. 6 and 7,
corresponding to Figs, 3 and 4, respectively.  Here we note that
Eq. (12) indicates that $G_F(k)$ is a function of both $k^-$ and
$k^+$. Therefore, the forward propagator of gluon has two
components---along $y^+$ and $y^-$---even in the collinear limit.
An exception is that the gluon propagator connects with the hard-
and soft-parts of the amplitude. The factorization scheme [12]
considers that soft gluons attach to this propagator at the
eikonal approximation. In this case, the largest ``+" component
$p^+$ of the three-gluon vertex in the numerator shall cancel
against a corresponding term in the denominator of the gluon
propagator and the resulting $G_F$ only contains its
$y^+$-component.

       Similarly, we can draw the twist-6 amplitudes. There are many different TOPT
diagrams. Some typical diagrams of them are shown in Figs. 8 and 9
for the transverse and longitudinal coefficient functions,
respectively. Figures 10 and 11 are corresponding geometric TOPT
diagrams. It is really difficult to judge which diagrams give
leading contributions to the nuclear size-dependent effect in
Figs. 3, 4, 8 and 9. However, we shall show that the leading
corrections in the corresponding geometric diagrams only come from
a few of diagrams (Figs. 6a, 10a and 10b), which contain maximum
number of colorless gluon pair, and such corrections can be simply
resummed.

\newpage
\begin{center}
\section{Normalization condition}
\end{center}

        Any calculation of the DIS cross section needs a factorization scheme
to separate the calculable partonic part from the nonperturbative
matrix element. In such a scheme, the propagators with the spinor
trace or the Lorentz indices linking two separated parts are
decomposed. At the same time, an arbitrary constant may sneak into
the undefined correlation function in this technique. Therefore,
at first we determine the normalization constant in the definition
of the correlation function. For this reason, we begin with the
naive parton model, which defines the quark distribution functions
$f_{q_i}(x_B)$. According to Fig. 12, the transverse structure
function $F_T$ at the leading order takes the form,
\begin{eqnarray}
F_T(x_B)&=&2e_T^{\mu\nu}W^{\mu\nu}\nonumber\\
&=&\sum_i e_i^2\int dx\frac{1}{2}Tr\left
[xP^+\gamma\cdot\overline{n}\gamma^{\mu}
\frac{Q^2}{2x_BP^+}\gamma\cdot
n\gamma^{\nu}\right]d^{\mu\nu}_{\perp}
\delta(x-x_B)\frac{x_B}{Q^2}T_{2q}(x)\nonumber \\
&=&\sum_i 2 e^2_ix_BT_{2q}(x_B)\equiv \sum_i e^2_if_{q_i}(x_B),
\end{eqnarray}
where we replace the propagator $(k'^2+i\epsilon)^{-1}$ with
$\delta(k'^2)= x_B\delta(x-x_B)/Q^2$ due to the on-shell condition
of the parton model. Thus, we have $f_{q_i}(x)=2xT_{2q}(x)$, where

\begin{equation}
T_{2q}(x)=\int\frac{P^+dy^-}{2\pi}e^{ixP^+y^-} <N\vert
\overline{\Psi}_i(0)\frac{\gamma^+}{2P^+}\Psi_i(y^-)\vert N>.
\end{equation}

        The normalization constant in the gluon distribution
function can be determined by using the contribution of gluons to
$F_T(x_B,Q^2)$ at the leading logarithmic approximation (Fig. 13).
In the TOPT form, we have

\begin{eqnarray}
F_T^{LLA}(x_B,Q^2)&=&d^{\mu\nu}_{\perp}W^{\mu\nu}\nonumber\\
&=&\sum_i e_i^2\int \frac{d^3\vec{k_1}}{(2\pi)^3}
<\frac{1}{2}>_{colour}4\pi\alpha_s\frac{E_p}{E_{\hat{k}_S}}
\frac{1}{8E_{\hat{k}_S}E_pE_{k_1}}\left (\frac{1}{E_{\hat{k}_S}+E_{k_1}-E_{p}}\right)^2\nonumber\\
& &\times \frac{1}{2}Tr[\gamma\cdot k_1\gamma^{\alpha}\gamma\cdot
\hat{k}_S\gamma^{\beta}] d^{\alpha\beta}_{\perp}
Tr[\gamma^{\mu}\gamma\cdot k_2\gamma^{\nu}\gamma\cdot
\hat{k}_S]d^{\mu\nu}_{\perp}
\delta(z_S-\frac{x_B}{y})\frac{x_B}{Q^2}T_{2g}\nonumber\\
&\simeq &\sum_i e^2_i\int \frac{dk^2_{\perp}}{k^2_{\perp}}dz_s
\frac{\alpha_s}{2\pi}[(\frac{x_B}{y})^2+(1-(\frac{x_B}{y})^2)]f_g(y),\nonumber\\
\end{eqnarray}
where the momenta of partons in Fig.13 are parameterized as
\begin{eqnarray}
p&=&\left [yP^+,\b{0},yP^+\right ],\nonumber\\
\hat{k}_S&=&\left [xP^++\frac{k^2_{\perp}}{2xP^+},k_{\perp},xP^+
\right ], \nonumber\\
k_1&=&\left
[(1-x)P^++\frac{k^2_{\perp}}{2(1-x)P^+},-k_{\perp},(1-x)P^+
\right].\nonumber\\
\end{eqnarray}

    The definition of $T_{2g}(x)$ in Eq. (18) is

\begin{eqnarray}
T_{2g}(x)=\int\frac{P^+dy^-}{2\pi}e^{ixP^+y^-} <N\vert
A^{+\alpha}(0)A^{+\beta} (y^-)\vert N>d^{\alpha\beta}_{\perp},\nonumber\\
\end{eqnarray} or

\begin{eqnarray} T_{2g}(x)=
\int\frac{P^+dy^-}{2\pi}e^{ixP^+y^-} <N\vert
F^{+\alpha}(0)F^{+\beta} (y^-)\vert
N>(\frac{1}{xP^+})^2d^{\alpha\beta}_{\perp} ,\nonumber\\
\end{eqnarray}
where $A^{+{\alpha}}\rightarrow F^{\alpha\beta}n^{\beta}/(xP^+)$
and $F^{\alpha\beta}$ is the gluon field strength. The second
definition has the gauge invariant form and presents singularity
when $x\rightarrow 0$, which implies the exchange of soft gluon
between the nonperturbative matrix and perturbative coefficient
function. However, all those soft gluons should be absorbed into
the definition of parton distribution functions (PDFs) in the
collinear factorization scheme. Therefore, the factor $1/x$ in Eq.
(21) should not give rise to any observable effects in the
perturbation part of the amplitude. For avoiding misunderstanding,
it is different from Ref. [6], we use the gluon distribution
function

\begin{eqnarray}
f_g(x)=\int\frac{P^+dy^-}{2\pi}e^{ixP^+y^-} <N\vert
A^{+\alpha}(0)A^{+\beta} (y^-)\vert N>d^{\alpha\beta}_{\perp}.
\end{eqnarray}

\newpage
\begin{center}
\section{Geometric nuclear enhancement dynamics}
\end{center}

   The nuclear target provides a special ``laboratory" for the resummation of
the higher twist effects.  The contributions of the nuclear
effects to the inclusive deep inelastic cross section are governed
by the hadronic tensor. In the factorization scheme, the operator
product in the hadronic tensor is expanded on the twist, where the
higher twist amplitudes include the contributions of the
multi-partonic configurations in the target. The contributions of
the partons from different nucleons may give rise to a visible
nuclear dependence in the structure functions.

     The nuclear effects in the structure functions originate from
the correlations of partons among the bound nucleons. The parton
correlation function is a generalization of the parton density
beyond the leading twist. A general correlation function for two
quarks and two gluons is defined as

\begin{eqnarray}
T_{2q2g}&=&\int\frac{P^+dy^-}{2\pi}\frac{P^+dy^-_1}{2\pi}\frac{P^+dy_2^-}{2\pi}
e^{ix_1P^+y^-}e^{i(x-x_1)P^+y_1^-}e^{-i(x-x_2)P^+y_2^-}\nonumber\\
&&\times\frac{1}{2P^+}\frac{1}{A}<A\vert\overline{\Psi}(0^-)
\gamma^+A^{+\alpha}(y_1^-)A^{+\beta}(y_2^-)\Psi(y^-) \vert
A>d^{\alpha\beta}_{\perp}.
\end{eqnarray}

    We always compare the structure function per bounded
nucleon with that of a free nucleon to present the nuclear effects.
Therefore, we do not expect a large nuclear dependence in the leading
twist process, since
\begin{eqnarray}
T_{2q}&=&\int\frac{P^+dy^-}{2\pi}e^{ixP^+y^-}\frac{1}{2P^+}
\frac{1}{A}<A\vert\overline{\Psi}(0)\gamma^+\Psi(y^-)\vert A>\nonumber\\
&\simeq& \int\frac{P^+dy^-}{2\pi}e^{ixP^+y^-}\frac{1}{2P^+}
<N\vert\overline{\Psi}(0)\gamma^+\Psi(y^-)\vert N>,
\end{eqnarray}
where $\vert A>$ and $\vert N>$ are the state vectors for nucleus
and free nucleon, respectively.

        In the general case, the exponential oscillations in the higher twist
matrix element destroy any nuclear size enhancement that could
come from the $y^-$ integration [7,13]. The contributions of the
integrations with respect to  $y^-_1$ and $y^-_2$ to Eq. (23) are
namely from the gluonic fields neighboring the targeted quarks. In
this case, we assume that

\begin{eqnarray}
\int\frac{P^+dy^-}{2\pi}\frac{P^+dy^-_1}{2\pi}\frac{P^+dy_2^-}{2\pi}
e^{ix_1P^+y^-}e^{i(x-x_1)P^+y_1^-}e^{-i(x-x_2)P^+y_2^-}
\frac{1}{2P^+}\nonumber\\
\times\frac{1}{A}<A\vert\overline{\Psi}(0^-)
\gamma^+A^{+\alpha}(y_1^-)A^{+\beta}(y_2^-)\Psi(y^-)\vert A>
d^{\alpha\beta}_{\perp}\nonumber\\
\simeq
\int\frac{P^+dy^-}{2\pi}\frac{P^+dy^-_1}{2\pi}\frac{P^+dy_2^-}{2\pi}
e^{ix_1P^+y^-}e^{i(x-x_1)P^+y_1^-}e^{-i(x-x_2)P^+y_2^-}
\frac{1}{2P^+}\nonumber\\
\times<N\vert\overline{\Psi}(0^-)
\gamma^+A^{+\alpha}(y_1^-)A^{+\beta}(y_2^-)\Psi(y^-)\vert N>
d^{\alpha\beta}_{\perp}.
\end{eqnarray}

    Therefore, we do not expect a nuclear size effect in Eq. (25).
However, some TOPT-diagrams for $F_T(x_B,Q^2)$ at twist-4 (Fig.
14) show that $y^-_1=y^-_2$. We further consider the case where
two gluonic operators consist a colorless cluster, which belongs
to the same extra nucleon. In this case there is no strong
correlation between the struck nucleon and the extra nucleon. The
momentum exchange among the bounded nucleons can be neglected in
comparison with a lager $P^+$-scale. Thus, we assume that
$x_1\simeq x_2$ in Eq. (25). Now all nucleons listing on the
integral path can contribute their gluons to the double scattering
in Fig. 14 without distortion.  Thus, we can write

\begin{eqnarray}
T_{2q2g}&=&\int\frac{P^+dy^-}{2\pi}\frac{P^+dy^-_1}{2\pi}
e^{ix_1P^+y^-}\frac{1}{A}\frac{1}{2P^+}\eta_{\perp}\nonumber\\
&&\times<A\vert\overline{\Psi}(0^-) \gamma^+\Psi(y^-)\vert A>
<A\vert A^{+\alpha}(y_1^-)A^{+\beta}(y_1^-)\vert A>
d^{\alpha\beta}_{\perp}\nonumber\\
&\simeq&\int\frac{P^+dy^-}{2\pi}\frac{P^+dy^-_1}{2\pi}
e^{ix_1P^+y^-}\frac{1}{2P^+}\eta_{\perp}<N\vert\overline{\Psi}(0^-)
\gamma^+\Psi(y^-)\vert N>\nonumber\\
&&\times<A\vert A^{+\alpha}(y_1^-)A^{+\beta}(y_1^-)\vert A>
d^{\alpha\beta}_{\perp}\nonumber\\
&=&\eta_{\perp}\eta_{\parallel}T_{2q}\Theta\nonumber\\
&=&\frac{A^{1/3}-1}{R^2}T_{2q}\Theta,
\end{eqnarray}
where
\begin{eqnarray}
\Theta\equiv\int\frac{P^+dy^-_1}{2\pi} <N\vert
A^{+\alpha}(y_1^-)A^{+\beta}(y_1^-)\vert
N>d^{\alpha\beta}_{\perp};
\end{eqnarray}
The factors $\eta_{\perp}$ and $\eta_{\parallel}$ are the
transverse and longitudinal correlation functions, respectively.
$Q^2\eta_{\perp}=Q^2/R^2$ can be schematically explained as the
overlap probability of two partons, where $1/Q$ is the scale of a
parton at momentum transfer $Q^2$ and $R$ is the maximum
correlation length of two partons. Usually, $R$ is regarded as an
effective size of nucleon.  The nuclear size effect occurs when
the scattered quark exceeds the longitudinal size of a nucleon,
and it reaches a maximum at $x_B=1/(2R_Am_N)$, where $R_A$ is
nuclear radius and $m_N$ the nucleon mass.

  Equation (26) shows a nuclear enhancement mechanism in the
higher twist processes, i.e, the rescattering of the outgoing
quark on the extra nucleons via the exchange of the contacting
gluon-pair can give rise to a visible $A^{1/3}$-dependence of the
matrix element. We regard a pair of gluons, which are contacted in
the $y^-$-space as the effective particle ($\Theta$ particle).

    From the geometric pictures in Figs. 6 and 7,
a straightforward prediction is that the leading contributions to
the nuclear size effect in the twist-4 structure functions are
from Fig. 6a and its conjugate diagram. However, the oscillating
factor can not be suppressed since $y^-_1\ne y^-_2$, although a
quark pair in Fig. 6c can come from an extra nucleon and forms a
colorless cluster. The contributions of Fig. 6b, 6d and 6e are
also excluded since there is no contacting gluon pair, which can
freely move along the longest path on the $y^-$-space. Thus, the
leading contributions of the geometric effect at twist-4 are only
from Figs. 6a and its conjugate diagram, with all the rest
diagrams being sub-leading and negligible in the resummation.

    The leading diagrams in twist-6 amplitudes are Figs. 10a, 10b and
their conjugates, which lead to the nuclear size effect $\sim
A^{2/3}$. On the other hand, the quark-gluon pairs in Figs.
10c-10e are colored and they are strongly correlated with the
other nucleons. Thus, we can not set $x_1=x_2$ in these diagrams
to get the leading nuclear size effect. The integrated path of a
contacting gluon-pair, which is connected with other nucleon via a
gluon in Figs. 10f, is shorter than that in Fig. 10a. Therefore,
we conclude that the contributions of Figs. 10c-10f, 11a-11d (and
other diagrams, which have not dawn in Figs. 10 and 11) are $\le
A^{2/3}$, and they are excluded in counting the leading effect. In
consequence, we need resum only the leading contributions from
such CLTOPT diagrams, where a maximum number of contracting
gluon-pairs may freely move along $y^-_1$ in a larger nucleus.
According to this rule, the contributions of the longitudinal
structure functions are excluded in our resummation. Therefore,
the geometry of the CLTOPT provides an ideal tool to choose the
leading diagrams in the resummation of the geometric nuclear
effect.

\newpage
\begin{center}

\section{Geometric nuclear size effect in structure functions}
\end{center}

    We calculate the contributions of higher twist amplitudes
to the structure functions in this Section. We begin with the
process shown in Fig. 14a. According to the results of the above
section, this process provides a leading nuclear effect to the
transverse structure function. Using the projection operator and
TOPT rules, the contributions of Fig. 14a to the $F_T(x_B,Q^2)$ at
$x_1\rightarrow x_2$ are
\begin{eqnarray}
\Delta_a F_T(x_B,Q^2)&=&2e^{\mu\nu}W^{\mu\nu}\nonumber\\
&=&\sum_i 4\pi e^2_i\alpha_s\int dx_1dx_2<\frac{1}{2}>_{colour}
d^{\mu\nu}_{\perp}\nonumber\\
&&\times \frac{1}{4}Tr\left[x_2P^+\gamma\cdot
\overline{n}\gamma^{\mu}\frac{\gamma\cdot n}
{2(x_1-x_B)P^+}\gamma^{\alpha}\gamma\cdot
n\frac{x_BP^+}{Q^2}\gamma^{\beta}
\frac{Q^2}{2x_BP^+}\gamma\cdot n\gamma^{\nu}\right]\nonumber\\
&&\times d^{\alpha\beta}_{\perp}
\delta(x_2-x_B)\frac{x_B}{Q^2}T_{2q2g}\nonumber\\
&=&\sum_i e^2_i\int
dx_1dx_2\frac{2\pi\alpha_s}{Q^2R^2}\frac{x_B\delta(x_2-x_B)}
{x_1-x_B}f_{q_i}(x_1,Q^2)(A^{1/3}-1)\Theta.
\end{eqnarray}

        Similarly, we have the contributions of Fig. 14b to $F_T$
at $x_2\rightarrow x_1$

\begin{eqnarray}
\Delta_bF_T(x_B,Q^2) =\sum_i e^2_i\int
dx_1dx_2\frac{2\pi\alpha_s}{Q^2R^2}
\frac{x_B\delta(x_1-x_B)}{x_2-x_B}f_{q_i}(x_1,Q^2)(A^{1/3}-1)\Theta.
\end{eqnarray}
Summing the contributions of Figs. 14a and 14b in the limit
$x_1\sim x_2$ (see Eq. (A.1) in Appendix A) we get

\begin{eqnarray}
\Delta F_T(x_B,Q^2) =\sum_i
e^2_i\frac{2\pi\alpha_s}{Q^2R^2}(A^{1/3}-1)\Theta
x_B\frac{d}{dx_B}f_{q_i}(x_B,Q^2).
\end{eqnarray}
Here we point out that the results of hard part using the TOPT
coincides with that using the covariant perturbation theory, which
was given in [14].

        We emphasize that the contributions of the geometric
nuclear effect to $F_T^{twist-4}(x_B,Q^2)$ are always negative at
small $x_B$ due to the shape of the quark number density as a
function $x_B$.  Equation (30) means that the double rescattering
of an outgoing quark via an effective $\Theta$ particle, which
consists of a pair of contacting gluons, reduces the magnitudes of
the transverse structure function.

        Now we consider the contributions of the higher twist amplitudes and
sum over them. We calculate the processes in Fig. 15, where the
cut line divides the initial gluons into two parts, each contains
an even number of gluons. For simplicity, we indicate the $\Theta$
particle as the dashed line. We have shown that these amplitudes
provide maximum nuclear size effects in the structure functions at
the twist-6 level. The contributions of Fig. 15a to $F_T(x_B,Q^2)$
at $x_1\rightarrow x_3$ is

\begin{eqnarray}
\Delta_a F_{T,twist-6}(x_B,Q^2) =\sum_i e^2_i\int
dx_1dx_2dx_3(\frac{2\pi\alpha_s}{Q^2R^2})^2
\frac{x_B^2\delta(x_3-x_B)} {(x_1-x_B)(x_2-x_B)}2x_3T_{2q4g},
\end{eqnarray}
where the matrix element is
\begin{eqnarray}
&&2x_3T_{2q4g}(x_1,Q^2)\nonumber\\
&=&2x_3\int\frac{P^+dy^-}{2\pi}\frac{P^+dy^-_1}{2\pi}
e^{ix_1P^+y^-}\frac{1}{A}\frac{1}{2P^+}\nonumber\\
&&\times <A\vert\overline{\Psi_i}(0^-)
\gamma^+A^{+\alpha_1}(y_1^-)A^{+\beta_1}(y_1^-)
\gamma^+A^{+\alpha_2}(y_2^-)A^{+\beta_2}(y_2^-) \Psi_i(y^-)\vert
A>d^{\alpha_1\beta_1}_{\perp}d^{\alpha_2\beta_2}_{\perp}
\nonumber\\
&=&(\frac{A^{1/3}-1}{R^2})^2{\Theta}^2f_{q_i}(x_B).
\end{eqnarray}
Therefore, we obtain
\begin{eqnarray}
&&\Delta_a F_{T,twist-6}(x_B,Q^2)\nonumber\\
&=& \sum_i e^2_i\int
dx_1dx_2dx_3(\frac{2\pi\alpha_s}{Q^2R^2})^2 (A^{1/3}-1)^2
\Theta^2\frac{x_B^2\delta(x_3-x_B)}
{(x_1-x_B)(x_2-x_B)}f_{q_i}(x_1,Q^2).
\end{eqnarray}
Summing the contributions of three cut diagrams in Fig. 15 at the
limit $x_1\rightarrow x_3$, we get (see Eq. (A.2) in Appendix A)
\begin{eqnarray}
\Delta F_{T,twist-6}(x_B,Q^2) =\sum_i
e^2_i(\frac{2\pi\alpha_s}{Q^2R^2})^2
(A^{1/3}-1)^2\frac{1}{2!}\Theta^2x_B^2\frac{d^2}{dx_B^2}f_{q_i}(x_B,Q^2).
\end{eqnarray}
Similarly, we obtain the contributions from the twist-$(2n+2)$
amplitudes to $F_T(x_B,Q^2)$, which contains the exchange of
$n$-$\Theta$ particles (see Eq. (A.3) in Appendix A). The result
is
\begin{eqnarray}
\Delta F_{T,twist-(2n+2)}(x_B,Q^2) =\sum_i
e^2_i(\frac{2\pi\alpha_s}{Q^2R^2})^n
(A^{1/3}-1)^n\frac{1}{n!}\Theta^nx_B^n\frac{d^n}{dx_B^n}f_{q_i}(x_B,Q^2).
\end{eqnarray}

        Thus, we can easily write the sum of the contributions from
all higher twist amplitudes, which give a leading nuclear size
dependence of the transverse structure functions $F_T(x_B,Q^2)$,
\begin{eqnarray}
\Delta F_T(x_B,Q^2)&=&\sum_i
e^2_i\sum_{n=1}^N(\frac{2\pi\alpha_s}{Q^2R^2})^n
(A^{1/3}-1)^n\frac{1}{n!}\Theta^nx_B^n\frac{d^n}{dx_B^n}f_{q_i}(x_B,Q^2)\nonumber\\
&=&\sum_i e^2_i\left[\exp \left(\frac{2\pi\alpha_s}{Q^2R^2}
(A^{1/3}-1)\Theta
x_B\frac{d}{dx_B}\right)-1\right]f_{q_i}(x_B,Q^2).
\end{eqnarray}

\newpage
\begin{center} \section{Geometric nuclear size effect in evolution equation}
\end{center}

     In this section we discuss the contributions of the geometric higher twist effect
to the DGLAP equation. We still work in the collinear infinite
frame (see Eq. (2)). The coefficient functions with one quark loop
corrections and at $LLA(Q^2)$ can be modelled using either the
standard scattering picture (Fig. 11) at $p^+\gg q^-$ or the
dipole picture (Fig. 16) at $q^-\gg p^+$ (we call the latter frame
as the dipole frame). There are different versions of the dipole
model. A general dipole model for the DIS [15] is constructed in
the $k_T$-factorization scheme, where the transverse momenta of
the initial partons do not vanish and evolution dynamics is the
BFKL equation [16]. In this work we shall show that the dipole
picture can also be used in the collinear factorization scheme.

    As we have pointed out that the
geometric nuclear effect is exposed in the collinear part of the
outgoing quark interacting with the nuclear gluons, where the
transverse momenta of the partons are negligible. Therefore, the
geometric nuclear effect is hidden under the scattering picture
Fig. 13, since the contributions of the transverse momentum
flowing into the quark-gluon vertex in this picture is necessary
in the derivation of the evolution (splitting) kernel.

    Different from the scattering picture, in the dipole picture (Fig.
16) the interacting part of quark with gluon (the dashed box in
Fig. 16) can be separated from the splitting kernel using TOPT and
then we may take the collinear approximation. At first step, we
derive the DGLAP equation in the dipole model. The momenta of the
partons in Fig. 16 are parameterized as

\begin{eqnarray}
q=q^-n-\frac{Q^2}{2q^-}\overline{n},
\end{eqnarray}
i.e.,

\begin{eqnarray}
E_q&=&\frac{1}{\sqrt{2}}(q^++q^-)\simeq \sqrt {2}q^-\equiv {q'}^-,  \nonumber\\
q_z&=&\frac{1}{\sqrt{2}}(q^+-q^-)\simeq -\sqrt {2}q^-= -{q'}^-.
\end{eqnarray}
Thus, we have

\begin{eqnarray}
\hat{k}_D&\simeq& (z_D{q'}^-+\frac{k_{\perp}^2}{2z_D{q'}^-}, k_{\perp}, -z_D{q'}^-),\nonumber\\
k_2&\simeq& ((1-z_D){q'}^-+\frac{k_{\perp}^2}{2(1-z_D){q'}^-},
-k_{\perp}, -(1-z_D){q'}^-).
\end{eqnarray}

    On the other hand,
\begin{eqnarray}
p=(E_p, p_{\perp}, p_3)=(y_DP^+, 0, y_DP^+).
\end{eqnarray}

    The relevant coefficient function which contains the splitting function
at $LLA(Q^2)$ in the dipole picture is

\begin{eqnarray}
C_D^{twist-2}&=&\int
\frac{d^3\vec{k}_1}{(2\pi)^3}\frac{1}{8E_{\hat{k}_D}E_{k_2}E_q}
\left [\frac{1}{E_{\hat{k}_D}-E_{k_2}-E_q}\right ]^2
\frac{1}{2}\sum\vert \overline{M}^D(\gamma \rightarrow q\overline{q})\vert^2 \nonumber\\
&\times&
\frac{E_q}{E_{\hat{k}_D}}\frac{1}{4}\sum\overline{M}^D(qg\rightarrow
q)\frac{1}{z_DQ^2}
\delta(z_D-\frac{y_D}{x_B})\nonumber\\
&\equiv& \frac{\sum e^2_i}{8\pi^2}\int
\frac{dk^2_{\perp}}{k^2_{\perp}}\frac{dy_D}{y_D}\frac{1}{z_DQ^2}
\left [z_D^2+(1-z_D)^2\right]H^{twist-2},
\end{eqnarray}
where we introduce the probe function of the gluon

\begin{eqnarray}
H^{twist-2}&=&\frac{1}{4}g^2Tr\left[z_D\frac
{Q^2}{2q'^-}\gamma\cdot\overline{n}\gamma_{\alpha}z_Dq'^-\gamma\cdot
n\gamma_{\beta} \right]d^{\alpha\beta}_{\perp} \nonumber\\
&=&g^2z_D^2Q^2,
\end{eqnarray}
and used the on-shell condition

\begin{eqnarray}
\delta((p+z_Dq)^2)=\frac{1}{z_DQ^2}\delta(z_D-\frac{y_D}{x_B}).
\end{eqnarray}  In the calculations of $\sum\vert \overline{M}^D(\gamma \rightarrow
q\overline{q})\vert^2$, we regard the virtual photon as an
equivalent photon in the Weizs$\ddot{a}$cker-Williams
approximation [17] under condition $q^-\gg p^+$. In this case, we
consider the contributions of the transverse polarized photon and
define $z_D$. Note that the longitudinal coefficient function
doesn't contribute to the DGLAP dynamics.

     The splitting function is defined as
\begin{eqnarray}
\frac{C_D^{twist-2}}{2z_D\sum e^2_i}=\frac{\alpha_s}{2\pi}\int
\frac{dk^2_{\perp}}{k^2_{\perp}} \frac{dy_D}{y_D}P_{qg}^D(z_D),
\end{eqnarray}
i.e.,

\begin{eqnarray}
P^D_{qg}(z_D)=\frac{1}{2}[z_D^2+(1-z_D)^2].
\end{eqnarray}
Using the factorized splitting function, we can construct a DGLAP
equation in the dipole picture (see Appendix B and [18])

\begin{eqnarray}
\frac{d f_{q_i}(x_B,Q^2)}{d \ln
Q^2}=\frac{\alpha_s}{2\pi}\int_{x^2_B}^{x_B}\frac{dy_D}{y_D}P^D_{q_ig}(\frac{y_D}{x_B})f_g(y_D,Q^2),
\end{eqnarray}
where $x_B^2\leq y_D\leq x_B$.

     Comparing Eq. (46) with the DGLAP equation in the scattering picture, we have the following
relations between the scattering and dipole pictures due to the
different definitions about $z_D$ and $z_S$:

\begin{eqnarray}
P^S_{qg}(z_S)\leftrightarrow P^D_{qg}(z_D),
\end{eqnarray}
where $z_D=y_D/x_B$, $z_S=x_B/y_S$ and we use ``D" and ``S" to
indicate the quantities in the dipole and scattering pictures,
respectively.

    We emphasize that although both $y_D$ and $y_S$ are defined as
the momentum fractions of the target carried by the gluon, they
have different kinematical regions.  It is interesting to note
that a smaller value of $x_B$ corresponds to a smaller value of
$y_D$ in the dipole model since $x_B=y_D/z_D$; on the other hand,
in the scattering picture the gluons can take a larger momentum
fraction $y_S$ of proton-momentum even in the small $x_B$ region
since $z_S=x_B/y_S\leq 1$.

    Using Eq. (47), one can rewrite Eq. (46) as a standard DGLAP
equation

\begin{eqnarray}
\frac{d f_{q_i}(x_B,Q^2)}{d \ln
Q^2}=\frac{\alpha_s}{2\pi}\int_{x_B}^{1}\frac{dy_S}{y_S}P^S_{qg}(\frac{x_B}{y_S})f_g(y_S,Q^2),
\end{eqnarray}
where $x_B\leq y_S\leq 1$.

    In our dipole model the splitting kernel containing loop
transverse momentum $k_{\perp}$ is separated from the probe
function $H^{twist-2}$, in which we can take the collinear
approximation, i.e., neglecting the contributions of the loop
transverse momentum.

    Now we discuss the corrections of the geometric nuclear effect to the DGLAP equation
using the dipole picture.  The nuclear gluons can collinearly
attach to two quark legs in any ways. However, the leading
contributions to the geometric nuclear effect are from following
probe functions:

\begin{eqnarray}
H^{twist-4}=H^{twist-2}2\pi\alpha_s(A^{1/3}-1)\frac{x^3_B}{y_DQ^2R^2}\Theta\frac{d}{dy_D}f_g(y_D,Q^2),
\end{eqnarray}
at twist-4 (see Fig. 17), and

\begin{eqnarray}
H^{twist-6}=H^{twist-2}(2\pi\alpha_s)^2(A^{1/3}-1)^2\frac{x^5_B}{y_D^2Q^4R^4}\Theta^2\frac{1}{2!}
\frac{d^2}{dy_D^2}f_g(y_D,Q^2).
\end{eqnarray}
at twist-6 (see Fig. 18).

    We sum the leading contributions of the nuclear size-effect at all order of twist to the DGLAP equation, the result
is

\begin{eqnarray}
&&\frac{d f_{q_i}(x_B,Q^2)}{d \ln Q^2}\nonumber\\
&&=\frac{\alpha_s}{2\pi}\int_{x_B}^{1}\frac{dy_S}{y_S}P^S_{q_ig}(\frac{x_B}{y_S})
f_g(y_S,Q^2)\nonumber\\
&&+\frac{\alpha_s}{2\pi}\int_{x^2_B}^{x_B}\frac{dy_D}{y_D}P^D_{q_ig}(\frac{y_D}{x_B})\left
\{x_B\left[\exp\left(2\pi\alpha_s
\frac{x_B^2}{y_DQ^2R^2}(A^{1/3}-1)\Theta\frac{d}{dy_D'}\right)-1\right]f_g(y_D',Q^2)\right
\}_{y'_D=y_D}.\nonumber\\
\end{eqnarray}

    We don't need to consider the nuclear size-effect on the gluon legs since these diagrams are
excluded as the sub-leading contributions in our resummation.

\newpage
\begin{center} \section{Discussions and summary}
\end{center}

   One of our results, Eq. (36), is consistent with the Qiu-Vitev resummation
for $\Delta F_T(x_B,Q^2)$ in Ref. [6], only differing from an
unknown constant. The reason is  that these two approaches are
based on different dynamics. The derivation of the Qiu-Vitev
formula assumes that the leading contributions to the nuclear size
effect in the structure functions are arisen from the
singularities of the gluon correlation functions. This argument
does not indicate a lot of neglecting graphs, which give the
sub-leading contributions. In this respect, the geometry of the
CLTOPT improves the Qiu-Vitev derivation.

      It is interesting to compare Eq. (51)
with the Mueller-Glauber model [2]. These two models are written
in the dipole picture but using different factorization schemes.
Besides, the Glauber approach of the multiple scattering theory is
based on the eikonal approximation, which assumes that the
scattering on different nucleons are incoherent. Thus, the phase
shifts from each scattering can sum up to an exponential form. In
the TOPT-language, the eikonal approximation implies that all
propagators connecting with nuclear gluons take their forward
component. On the other hand, Eq. (51) provides a different
picture, where the backward propagator alternates with the forward
one when scattered quark passes through the nucleus.

     It is well known that there is an obvious nuclear screening effect
in the structure function $F_2^A$ at small $x_B$ [19]. As we have
pointed out that the nuclear size effect in the longitudinal
stricture function at leading order vanishes. The nuclear
screening effect at small $x_B$ region can be expressed in the
ratios of the structure functions as
\begin{eqnarray}
R(x_B,Q^2)&=&\frac{F_{A,2}(x_B,Q^2)}{F_{N,2}(x_B,Q^2)}\nonumber\\
&=& \frac{x_BF_{A,L}(x_B,Q^2)+x_BF_{A,T}(x_B,Q^2}{F_{N,2}(x_B,Q^2)}\nonumber\\
&\simeq& 1+\frac{x_B\Delta F_T(x_B,Q^2)}{F_{N,2}(x_B,Q^2)}.
\end{eqnarray}
From these results, one can fix the parameter $\Theta$.

    The nuclear PDFs are one of fundamental knowledge for understanding
the experimental data at the Relativistic Heavy Ion Collider
(RHIC) and Large Hadron Collider (LHC). The PDFs are calculable if
the nonperturbative input distributions and their QCD evolution
dynamics are known. Therefore, the corrections of nuclear
circumstances to the evolution equations and input distributions
become an active topic. The nuclear PDFs can be modelled using the
DGLAP equation from the relating DIS data [20]. The results show
that the input PDFs in the bound nucleon differ by the shadowing
factors from that in the free nucleon. However, this result does
not means that the corrections from the higher twist effects in
such researches are unimportant since the shadowing factors are
phenomenological. In fact, the nonlinear modification of the gluon
recombination to the DGLAP equation as a higher twist effect can't
be neglected even in DIS on the free nucleon [21]. As mentioned
already in works [6], the higher twist resummation may be of
importance in the extraction of the nuclear input PDFs. Our
results remind us of that, except the contributions of the gluon
recombination effect to the DGLAP equation, the geometric higher
twist effect in the nuclear input PDFs and QCD evolution equation
should be considered in extracting the nuclear parton
distributions from the measured DIS cross sections.

    As we have mentioned that the higher twist effects and their summation
may provide a valuable information for the multi-parton
correlations at long distance in the target.  The geometric
nuclear effect presents an interesting freedom in high energy
nuclear physics---the contacting gluon-pair, which is soft at the
hadronic level since it carries almost zero-momentum
($x_1-x_2\simeq 0$), however, the effect is really calculable in
the perturbative QCD.

     In summary, the higher twist effect in deep inelastic scattering on
the nuclear target are studied using time ordered perturbation
theory. We have shown that the collinear rescattering of the
outgoing quark on the extra nucleon via the contacting gluon-pair
dominates a nuclear size dependent effect. The leading
contributions of nuclear-enhanced effect to the structure function
and DGLAP evolution equation are resummed by using the geometric
properties of the higher twist amplitudes.

\noindent {\bf Acknowledgments}: This work was supported by
National Natural Science Foundations of China 10075020, 50193013,
and 10475028.

\newpage
\noindent {\bf Appendix A}:

    We list some algebraic identities in the derivations in Section 5.

In $F_{T, twist-4}$

\begin{eqnarray*}
&&\int dx_1dx_2\left [\frac{\delta(x_2-x_B)}{x_1-x_B}+\frac{\delta(x_1-x_B)}{x_2-x_B}\right ]
\delta(x_2-x_1-\Delta)\\
&&=-\int dx_1\left [\frac{\delta(x_1-x_B+\Delta)}{\Delta}-\frac{\delta(x_1-x_B)}{\Delta}\right ]\\
&&=-\int dx_1[\delta'(x_1-x_B)]\\
&&=\frac{d}{dx_B},~~~~~~~~~~~~~~~~~~~~~~~~~~~~~~~~~~~~~~~~~~~~~~~~~~~~(A.1)
\end{eqnarray*}
when $\Delta\rightarrow 0$.

In $F_{T, twist-6}$

\begin{eqnarray*}
&&\int dx_1dx_2dx_3\left [\frac{\delta(x_3-x_B)}{(x_1-x_B)(x_2-x_B)}+\frac{\delta(x_2-x_B)}{(x_1-x_B)
(x_3-x_B)}+\frac{\delta(x_1-x_B)}{(x_2-x_B)(x_3-x_B)}\right ]\\
&&\times \delta(x_2-x_1-\Delta)\delta(x_3-x_2-\Delta)\\
&&=\int dx_1dx_2dx_3\left
[\frac{\delta(x_3-x_B)}{2\Delta^2}-\frac{\delta(x_2-x_B)}{\Delta^2}
+\frac{\delta(x_1-x_B)}{2\Delta^2}\right ]\delta(x_2-x_1-\Delta)\delta(x_3-x_2-\Delta)\\
&&=-\int dx_1dx_2\frac{1}{2\Delta}[\delta'(x_1-x_B)-\delta'(x_2-x_B)]\delta(x_2-x_1-\Delta)\\
&&=-\int dx_1\frac{1}{2\Delta}[\delta'(x_1-x_B)-\delta'(x_1-x_B+\Delta)]\\
&&=\int dx_1\frac{1}{2}\delta''(x_1-x_B)\\
&&=\frac{1}{2!}\frac{d^2}{dx_B^2},~~~~~~~~~~~~~~~~~~~~~~~~~~~~~~~~~~~~~~~~~~~~~~~~~~~~~~~~~~~~~~~(A.2)
\end{eqnarray*}
when $\Delta\rightarrow 0$.

In $F_{T, twist-8}$

\begin{eqnarray*}
&&\int dx_1dx_2dx_3dx_4\left [\frac{\delta(x_4-x_B)}{(x_1-x_B)(x_2-x_B)(x_3-x_B)}+\frac{\delta(x_3-x_B)}{(x_1-x_B)
(x_2-x_B)(x_4-x_B)}\right.\\
&&\left.+\frac{\delta(x_2-x_B)}{(x_1-x_B)(x_3-x_B)(x_4-x_B)}+\frac{\delta(x_1-x_B)}{(x_2-x_B)(x_3-x_B)(x_4-x_B)}\right ]\\
&&\times \delta(x_2-x_1-\Delta)\delta(x_3-x_2-\Delta)\delta(x_4-x_3-\Delta)\\
&&=\int dx_1dx_2dx_3dx_4\left [-\frac{\delta(x_4-x_B)}{6\Delta^3}+\frac{\delta(x_3-x_B)}{2\Delta^3}
-\frac{\delta(x_2-x_B)}{2\Delta^3}+\frac{\delta(x_1-x_B)}{6\Delta^3}\right ]\\
&&\times\delta(x_2-x_1-\Delta)\delta(x_3-x_2-\Delta)
\delta(x_4-x_3-\Delta)\\
&&=-\int dx_1dx_2dx_3\frac{1}{6\Delta^2}[\delta'(x_3-x_B)-\delta'(x_2-x_B)-\delta'(x_2-x_B)+\delta'(x_1-x_B)]\\
&&\times\delta(x_2-x_1-\Delta)\delta(x_3-x_2-\Delta)\\
&&=-\int dx_1\frac{1}{6\Delta}[\delta''(x_1-x_B+\Delta)-\delta''(x_1-x_B)]\\
&&=-\int dx_1\frac{1}{6}\delta'''(x_1-x_B)\\
&&=\frac{1}{3!}\frac{d^3}{dx_B^3},~~~~~~~~~~~~~~~~~~~~~~~~~~~~~~~~~~~~~~~~~~~~~~~~~~~~~~~~~~~~~~~(A.3)
\end{eqnarray*}
when $\Delta\rightarrow 0$.

\newpage
\noindent {\bf Appendix B}:

    In this Appendix we derive a set of complete DGLAP evolution
equations in the dipole frame. At first step, we give the
definitions of the parton distributions in this frame. The QCD
evolution equations of the PDFs are irrelevant to the probe,
therefor, we may use a virtual gluon or a virtual quark, which
connects with an extra color source ``S", to replace the virtual
photon in DIS. Thus, the first order of the DIS amplitudes in the
dipole frame, can be expressed as the four graphs shown in Fig.
19.

    We take Fig. 19a as an example. The cross section of a virtual
gluon scattering off a proton in the dipole frame and in the
collinear factorization scheme is

$$d\sigma_a(qP\rightarrow k_1X)=\int
dyf_g(y)d\sigma_a(qp\rightarrow k_1). \eqno(B.1)$$ For
convenience, we neglect the symbol``D" for the dipole
picture. According to the parton model,\\

$$d\sigma_a(qp\rightarrow k_1)=C_a\delta(y-x_B), \eqno(B.2)$$
where $C_a$ is a function describing the probe vertex. In the
above equation and below, we will use the momenta of the partons
to represent the particles entering cross sections and matrices.
The gluon distribution in the proton at the first order QCD can
now be defined as

$$f_g(x_B)=\frac{1}{C_a}d\sigma_a(qP\rightarrow k_1X).
\eqno(B.3)$$

Similarly, we have three more definitions corresponding to Fig.
19b-19d.

$$f_g(x_B)=\frac{1}{C_b}d\sigma_b(qP\rightarrow k_1X),
\eqno(B.4)$$

$$f_q(x_B)=\frac{1}{C_c}d\sigma_c(qP\rightarrow k_1X),
\eqno(B.5)$$ and

$$f_q(x_B)=\frac{1}{C_d}d\sigma_d(qP\rightarrow k_1X).
\eqno(B.6)$$

    On the other hand, we can write the formulae for the four
cross sections of Fig. 19 in the TOPT form as

$$d\sigma_{a,b}(qP\rightarrow k_1X)$$
$$=\frac{E_p}{E_P}\vert\overline{M}_g(P\rightarrow pX)\vert^2 \left
[\frac{1}{E_P-E_p-E_X}\right ]^2
\left[\frac{1}{2E_p}\right]^2\prod_X\frac{d^3k_X}{(2\pi)^32E_X}$$
$$\times \frac{1}{8E_pE_q}\overline{M}_{a,b}(p\rightarrow qk_1\rightarrow p)(2\pi)^4
\delta^4(p-q-k_1)\frac{d^3k_1}{(2\pi)^32E_{k_1}}, \eqno(B.7)$$ and

$$d\sigma_{c,d}(qP\rightarrow k_1X)$$
$$=\frac{E_p}{E_P}\vert\overline{M}_q(P\rightarrow pX)\vert^2 \left
[\frac{1}{E_P-E_p-E_X}\right ]^2
\left[\frac{1}{2E_p}\right]^2\prod_X\frac{d^3k_X}{(2\pi)^32E_X}$$
$$\times \frac{1}{8E_pE_q}\overline{M}_{c,d}(p\rightarrow qk_1\rightarrow p)(2\pi)^4
\delta^4(p-q-k_1)\frac{d^3k_1}{(2\pi)^32E_{k_1}}. \eqno(B.8)$$
Comparing Eqs. (B.7) and (B.8) with (B.1), we get the definitions
of the quark and gluon distributions

$$f_g(y)dy=\frac{E_p}{E_P}\vert\overline{M}_g(P\rightarrow pX)\vert^2 \left
[\frac{1}{E_P-E_p-E_X}\right ]^2
\left[\frac{1}{2E_p}\right]^2\prod_X\frac{d^3k_X}{(2\pi)^32E_X},
\eqno(B.9)$$

$$f_q(y)dy=\frac{E_p}{E_P}\vert\overline{M}_q(P\rightarrow pX)\vert^2 \left
[\frac{1}{E_P-E_p-E_X}\right ]^2
\left[\frac{1}{2E_p}\right]^2\prod_X\frac{d^3k_X}{(2\pi)^32E_X},
\eqno(B.10)$$ and the probe vertices

$$\frac{1}{C_j}d\sigma_j(qp\rightarrow k_1)$$
$$=\frac{1}{C_j}\frac{1}{8E_pE_q}\overline{M}_j(p\rightarrow qk_1\rightarrow p)(2\pi)^4
\delta^4(p-q-k_1)\frac{d^3k_1}{(2\pi)^32E_{k_1}}$$
$$=\delta(y-x_B) \eqno(B.11)$$ in the TOPT form, where $j=a, b, c$ and $d$.

    When the scale $Q^2$ is increased
by a small amount due to the radiation of a parton, i.e.,
$Q^2\rightarrow Q^2+\Delta Q^2$, the corresponding change of the
PDFs, for example in Fig. 20a' is

$$df_g(x_B)$$
$$=\frac{1}{C_a}d\sigma_{a'}(qP\rightarrow k_1k_2X)$$
$$=\frac{E_p}{E_P}\vert\overline{M}_g(P\rightarrow pX)\vert^2 \left
[\frac{1}{E_P-E_p-E_X}\right ]^2
\left[\frac{1}{2E_p}\right]^2\prod_X\frac{d^3k_X}{(2\pi)^32E_X}H_a(qp\rightarrow
qp)$$
$$=\int dyf_q(y)H_a(qp\rightarrow qp), \eqno(B.12)$$
where

$$H_a(qp\rightarrow qp)$$
$$=\frac{1}{C_a}d\sigma_a(p\rightarrow z_Dqk_1\rightarrow p)\frac{E_{\hat{k}}}{E_q}
\vert\overline{M}(q\rightarrow \hat{k}k_2)\vert^2 \left
[\frac{1}{E_{\hat{k}}-E_{k_2}-E_q}\right ]^2
\left[\frac{1}{2E_{\hat{k}}}\right]^2\frac{d^3k_2}{(2\pi)^32E_{k_2}}$$
$$=\frac{1}{y}\delta(z_D-\frac{y}{x_B})\frac{dk^2_{\perp}}{k^2_{\perp}}\frac{\alpha_s}{2\pi}P_{gg}(z_D),
\eqno(B.13)$$ where $P_{gg}(z_D)$ is the splitting function for
$g\rightarrow gg$.  The momenta of the partons in Fig. 20 are
parameterized as Eqs. (39) and (40).  From Eqs. (B.12) and (B.13),
we get a DGLAP evolution equation

$$\frac{df_g(x_B,Q^2)}{d\ln Q^2}=\frac{\alpha_s}{2\pi}\int_{x_B^2}^{x_B}\frac{dy}{y}f_g(y,Q^2)P_{gg}(\frac{y}{x_B}),
\eqno(B.14)$$ and

$$P_{gg}(z_D)=2C_2(G)\left[\frac{1-z_D}{z_D}+\frac{z_D}{1-z_D}+z_D(1-z_D)\right].
\eqno(B.15) $$

    Along similar lines, we derived the evolution equations
corresponding to Figs. 20b'-20d'

$$\frac{df_{q_i}(x_B,Q^2)}{d\ln Q^2}=\frac{\alpha_s}{2\pi}\int_{x_B^2}^{x_B}\frac{dy}{y}f_g(y,Q^2)P_{q_ig}(\frac{y}{x_B}),
\eqno(B.16)$$

$$\frac{df_g(x_B,Q^2)}{d\ln Q^2}=\frac{\alpha_s}{2\pi}\sum_{i=1}^{2f}\int_{x_B^2}^{x_B}\frac{dy}{y}f_{q_i}(y,Q^2)P_{gq_i}(\frac{y}{x_B}),
\eqno(B.17)$$ and

$$\frac{df_{q_i}(x_B,Q^2)}{d\ln
Q^2}=\frac{\alpha_s}{2\pi}\int_{x_B^2}^{x_B}\frac{dy}{y}f_{q_j}(y,Q^2)P_{q_iq_j}(\frac{y}{x_B}).
\eqno(B.18)$$

    The splitting functions are

$$P_{q_ig}(z_D)=\frac{1}{2}\left[z_D^2+(1-z_D)^2\right],
\eqno(B.19) $$

$$P_{gq_i}(z_D)=C_2(R)\left[\frac{1+(1-z_D)^2}{z_D}\right],
\eqno(B.20) $$and

$$P_{q_iq_j}(z_D)=C_2(R)\left[\frac{1+z_D^2}{1-z_D}\right].
\eqno(B.21) $$ Note that two partons indicated $\hat{k}$ in Fig.
20 are on mass-shell in the TOPT framework, therefore, one can
extract the probe function $H^{twist-2}$ from the amplitude (Fig.
20b') and generalize it to the higher twist processes as in
Section 6.

    The probabilistic form of the splitting functions in Eqs. (B.15) and
(B.21) have $1/(1-z_D)$ singularities, which are caused by the
emission of a low-momentum gluon. Physically, the infrared
singularities can be cancelled when the virtual diagrams, for
example Fig. 21 are added to the real contributions. As we have
pointed out in Ref. [22], an advantage of TOPT is that it exposes
the relation between the real and virtual diagrams without an
actual calculation of the virtual diagrams. In fact, from the
TOPT-decomposition of the diagrams, we can easily get

$$\frac{\alpha_s}{2\pi}\frac{dk^2_T}{k^2_T}P_{gg}^{virtual}$$
$$=\frac{1}{2}\frac{1}{8E_{\hat{k}}E_{k_2}}\frac{1}{E_{\hat{k}}+E_{k_2}-E_q}
\frac{1}{E_q-E_{\hat{k}}-E_{k_2}}\vert\overline{M}(g\rightarrow
\hat{k}k_2)\vert^2\frac{d^3\vec{k}_2}{(2\pi)^3}$$
$$=-\frac{1}{2}\frac{\alpha_s}{2\pi}\frac{dk^2_T}{k^2_T}P_{gg}, \eqno(B.22)$$
where $d^3\vec{k}_2$ is the loop-momentum integral; the factor
$1/2$ can be explained as follows: only half of the probe-vertex
connects with the partonic matrix in the virtual diagrams. That is

$$\frac{\sqrt{f_g}d\sqrt{f_g}}{d\ln Q^2}=\frac{1}{2}
\frac{df_g}{d\ln Q^2}. \eqno(B.23)$$

    The contributions from others virtual diagrams can be derived by
similar method. In consequence, we get the evolution equations

$$\frac{df_{q_i}(x_B, Q^2)}{d\ln Q^2}$$
$$=\frac{\alpha_s}{2\pi}\int^{x_B}_{x_B^2}\frac{dy}{y}f_{q_j}(y,Q^2)P_{q_iq_j}\left(\frac{y}{x_B}\right)
-\frac{\alpha_s}{2\pi}f_{q_j}(x_B,Q^2)\int_{x_B}^1dz_DP_{q_iq_j}(z_D)$$
$$+\frac{\alpha_s}{2\pi}\int_{x_B^2}^{x_B}\frac{dy}{y}f_g(y,Q^2)P_{q_ig}\left(\frac{y}{x_B}\right),
\eqno(B.24)$$ and

$$\frac{df_g(x_B, Q^2)}{d\ln Q^2}$$
$$=\frac{\alpha_s}{2\pi}\int^{x_B}_{x_B^2}\frac{dy}{y}\sum_{i=1}^{2f}f_{q_i}(y,Q^2)P_{gq_i}\left(\frac{y}{x_B}\right)
-\frac{\alpha_s}{2\pi}\sum_{i=1}^{2f}f_{q_i}(x_B,Q^2)\int_{x_B}^1dz_DP_{gq_i}(z_D)$$
$$+\frac{\alpha_s}{2\pi}\int^{x_B}_{x_B^2}\frac{dy}{y}f_g(y,Q^2)P_{gg}\left(\frac{y}{x_B}\right)
-\frac{\alpha_s}{2\pi}f_g(x_B,Q^2)\int_{x_B}^1dz_DP_{gg}(z_D).
\eqno(B.25)$$

\newpage

Figure Captions

\noindent Fig. 1 Elemental vertices of QCD at the collinear limit.

\noindent Fig. 2 Elemental vertices of QED at the collinear limit.

\noindent Fig. 3 The twist-4 TOPT diagrams at lowest order for
$F_T(x_B,Q^2)$, the conjugant diagrams have not drawn.

\noindent Fig. 4 As similar to Fig. 3, but for $F_L(x_B,Q^2)$.

\noindent Fig. 5 An example of time-space picture corresponding to
Fig. 3a.

\noindent Fig. 6 Space-time pictures of the amplitudes
corresponding to Fig. 3.

\noindent Fig. 7 Space-time pictures of the amplitudes
corresponding to Fig. 4.

\noindent Fig. 8 A part of twist-6 TOPT diagrams at lowest order
for $F_T(x_B,Q^2)$, the conjugant diagrams have not drawn.

\noindent Fig. 9 A part of twist-6 TOPT diagrams at lowest order
for $F_L(x_B,Q^2)$, the conjugant diagrams have not drawn.

\noindent Fig. 10 Space-time pictures of the amplitudes
corresponding to Fig. 8.

\noindent Fig. 11 Space-time pictures of the amplitudes
corresponding to Fig. 9.

\noindent Fig. 12 The naive parton model, which defines the quark
distributions.

\noindent Fig. 13 The leading order contribution of gluon to
$F_T(x_B,Q^2)$ in the standard parton (scattering) model.

\noindent Fig. 14 The leading perturbative contribution to the
nuclear size-dependent of $F_T(x_B,Q^2)$ at twist-4, where an
outgoing quark rescattering on the extra nucleon via the
contacting colorless gluon-pair exchange.

\noindent Fig. 15 The leading diagrams at the twist-6 level, where
using the dashed line to denote a contacting colorless gluon pair.

\noindent Fig. 16  Schematic diagrams for the decompositions of a
DGLAP equation at the dipole frame in TOPT.

\noindent Fig. 17 Space-time picture for the probing function at
twist-4, which gives the leading contribution to the nuclear
size-dependent of the DGLAP equation.

\noindent Fig. 18 Similar to Fig. 17, but at twist-6.

\noindent Fig. 19 The definitions of the parton distributions in
the dipole frame. Where $S$ is the color source.

\noindent Fig. 20 The one-loop corrections to the parton
distributions in the dipole picture.

\noindent Fig. 21  The virtual diagrams corresponding to Fig. 20a.

\newpage
\begin{center}



\end{center}


\newpage
\begin{thebibliography}{99}

\bibitem{1} G. Sterman, An Introduction to Quantum Field Theory (Cambridge
Univ. Press, Cambridge, 1993).

\bibitem{2} A.H. Mueller, Nucl. Phys. B 335 (1990) 115, B 415 (1994) 375 (1994);
B 437 (1995) 107.

\bibitem{3} J. Jalilian-Marian, A. Kovner, L. McLerran, and H. Weigert,
Phys. Rev. D 55, 5414 ~1997!; J. Jalilian-Marian, A. Kovner, A.
Leonidov, and H. Weigert, Nucl. Phys. B504 (1997) 415; Phys. Rev.
D 59 (1999) 014014; H. Weigert, Nucl. Phys. A 703 (2002) 823; E.
Iancu, A. Leonidiv, and L. McLerran, ibid. A 692 (2001) 583 ;
Phys. Lett. B 510 (2001) 133.

\bibitem{4} A.H. Mueller, Phys. Lett. B 525 (2001) 243.

\bibitem{5} C. S. Lam, G. Mahlon and W. Zhu, Phys. rev. D 66 (2002) 074005.

\bibitem{6} J.W. Qiu and I. Vitev, Phys.Lett. B587 (2004) 52; arXiv:hep-ph/0309094 and 0405068.

\bibitem{7} X.F. Guo, J.W. Qiu and W. Zhu, Phys. Lett. B 523 (2001) 88.

\bibitem{8} M.D. Scadron, 1979. Advanced Quantum theory and Its Applicatiopns
Through Feynman Diagrams, eds W. Beigb$\ddot{o}$ck, R.P. Geroch, E.H. Lieb,
T. Regge and W. Thirring (Springer-Verlag, Berlin Heidelberg) p. 159.

\bibitem{9} W. Zhu, H.W. Xiong and J.H. Ruan, Phys. Rev. D60 (1999) 094006.

\bibitem{10} G. Altarelli and G. Parisi, Nucl. Phys. B126 (1977) 298.

\bibitem{11} V.N. Gribov and L.N. Lipatov, Sov. J. Nucl. Phys. 15 (1972) 438;
Yu.L. Dokshitzer, Sov. Phys. JETP. 46 (1977) 641

\bibitem{12} J.C. Collins, D.E. Soper and G. Sterman, 1989,  Perturbative Quantum
Chromodynamics, ed A.H. Mueller (World Scientific) p. 1.

\bibitem{13} J.W. Qiu and G. Sterman, Phys. Rev. D 49 (1994) 4493; Phys. Rev. D 50 (1994) 1951.

\bibitem{14}  X.F. Guo and J.W. Qiu, arXiv:hep-ph/9810548.

\bibitem{15} A. H. Mueller, Parton Saturation-An Overview, hep-ph/0111244,
Published in QCD Perspectives on Hot and Dense Matter, Eds. J.-P.
Blaizot and E. Iancu, NATO Science Series, Kluwer, (2002) and
references therein.

\bibitem{16} L.N. Lipatov, Sov. J. Nucl. Phys. 23 (1976) 338; V. S. Fadin, E.A. Kuraev
and L.N. Lipatov, Phys. Lett. B 60 (1975) 50; E.A. Kuraev, L.N.
Lipatov and V. S. Fadin, Sov. Phys. JETP 44 (1976) 443; E.A.
Kuraev, L.N. Lipatov and V. S. Fadin, Sov. Phys. JETP 45 (1977)
199; I. I. Balitsky and L.N. Lipatov, Sov. J. Nucl. Phys. 28
(1978) 822; I. I. Balitsky and L.N. Lipatov, JETP Lett. 30 (1979)
355.

\bibitem{17} von Weizs$\ddot{a}$cker, C.F., 1934. Z. Phys. 88, 612; Williams,
E.J., 1934. Phys. Rev. 45, 729; P. Kessler, Nuovo Comento 16
(1966) 809; V.N Baier, V.S. Fadin and V.A. Khoze,  Nucl. Phys. 65
(1973) 381; M.S. Chen and P. Zerwas, Phys. Rev. D12 (1975)187.

\bibitem{18} Z.Q. Shen and W. Zhu, Chinese Phys. Lett. 21 (2004) 1896.

\bibitem{19} For a review, see M. Arneodo, Phys. Rep. 240 (1994) 301, and references therein.

\bibitem{20} K.J. Eskola, H. Honkanen, V.J. Kolhinen, P.V. Ruuskane and C.A.
Salgado, J. Phys. G29 (2003) 1947; M. Hirai, S. Kumano, and T.H.
Nagai, Phys.Rev. C70 (2004) 044905.

\bibitem{21} K.J. Eskola, H. Honkanen, V.J. Kolhinen, Jianwei Qiu, and
C.A. Salgado, Nucl. Phys. B660 (2003) 211; W. Zhu, J.H. Ruan, J.F.
Yang and Z.Q. Shen, Phys. Rev. D68 (2003) 094015.

\bibitem{22} W. Zhu, Nucl.Phys. B551 (1999) 245; W. Zhu and J.H. Ruan, Nucl. Phys. B559 (1999)
378; W. Zhu, Z.Q. Shen and J.H. Ruan, Nucl. Phys. B692 (2004) 417.
\end{thebibliography}
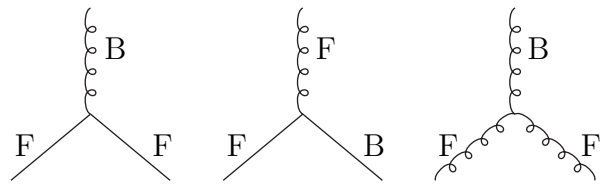
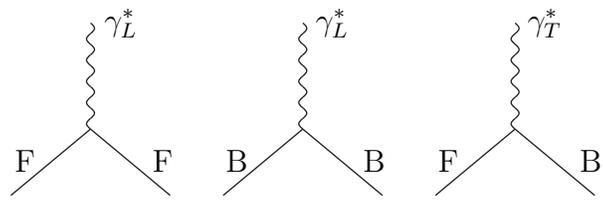
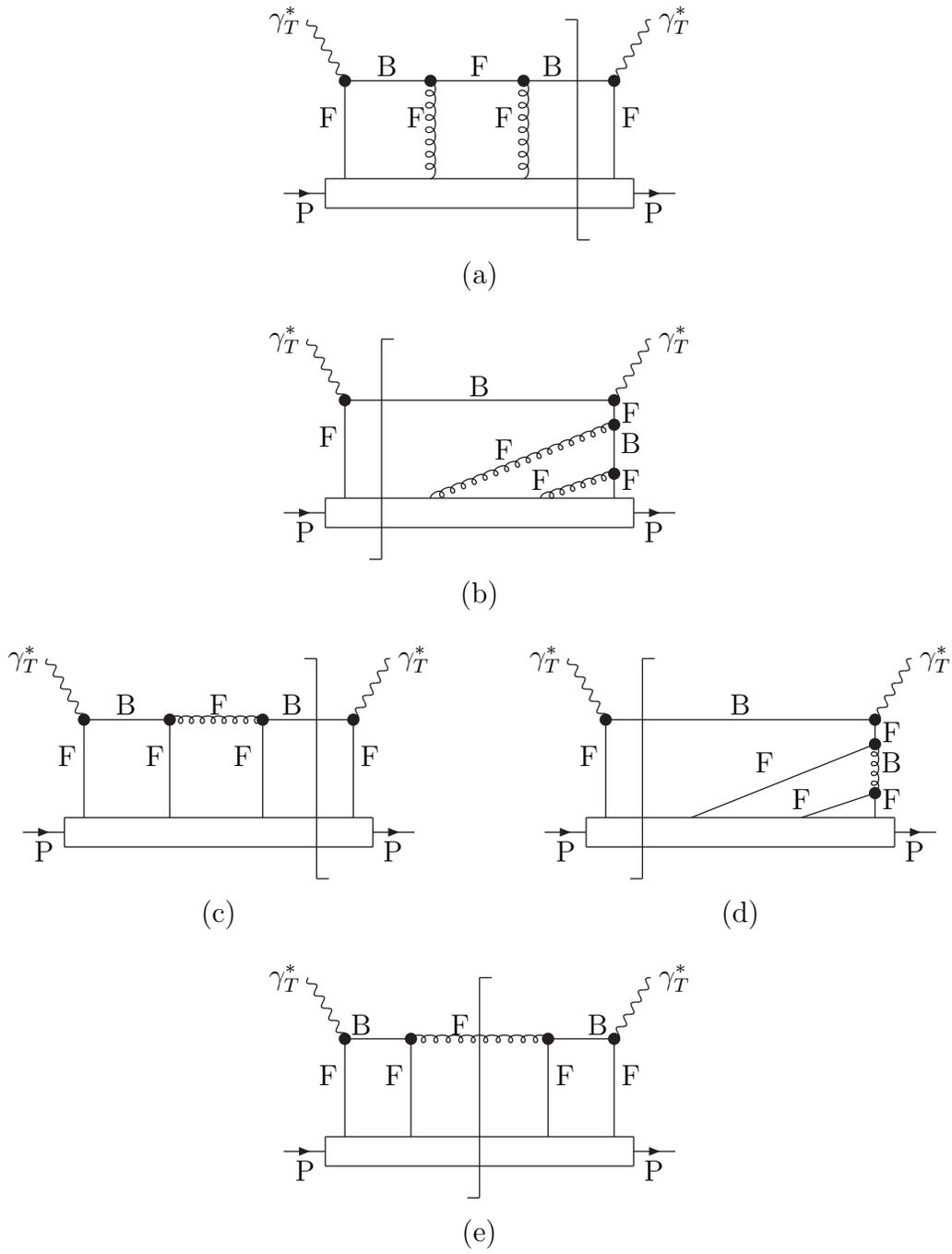
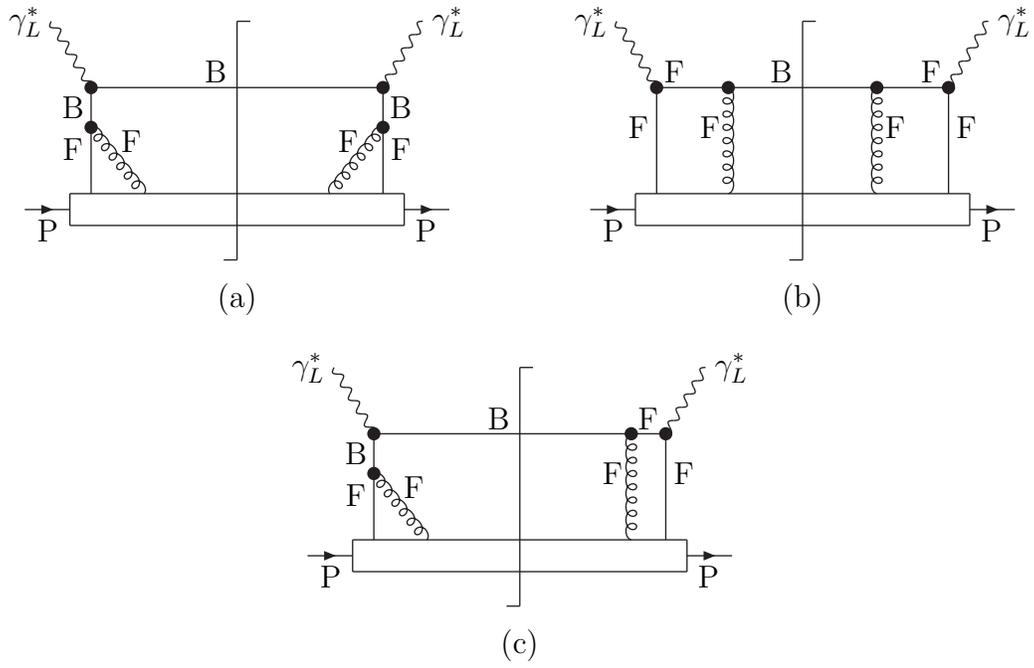
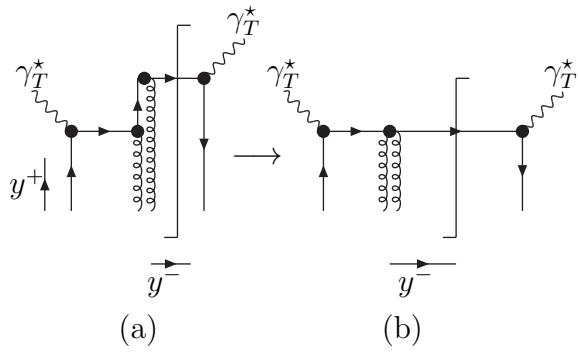
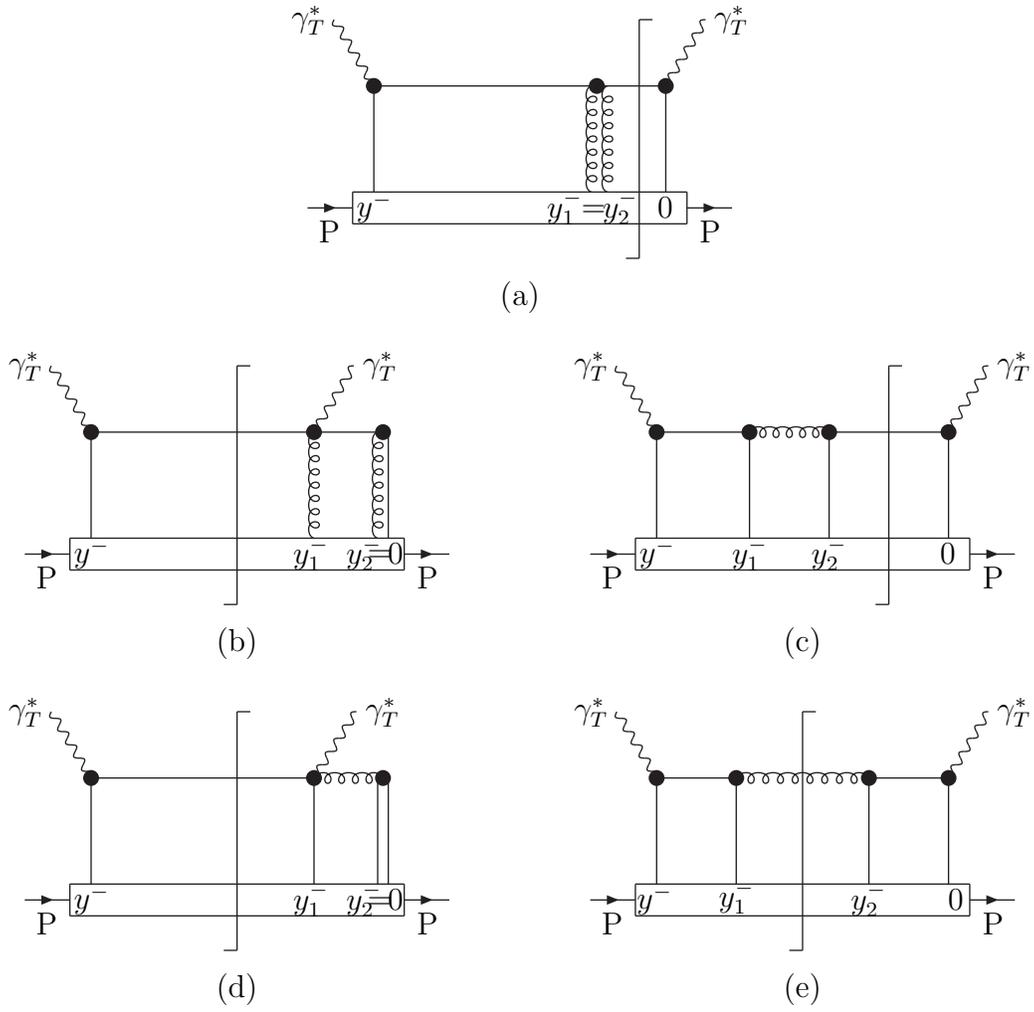
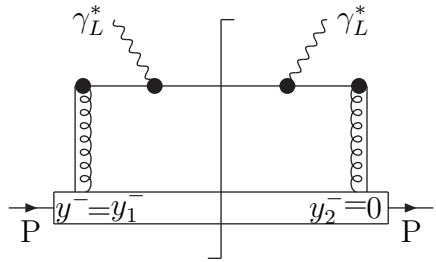
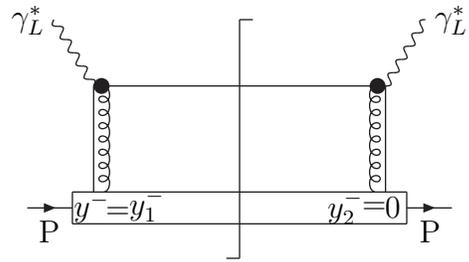
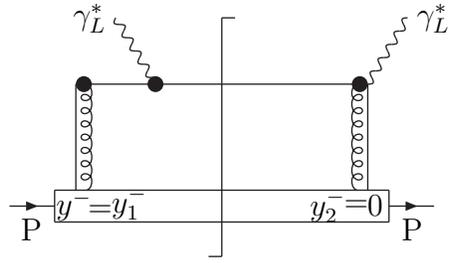
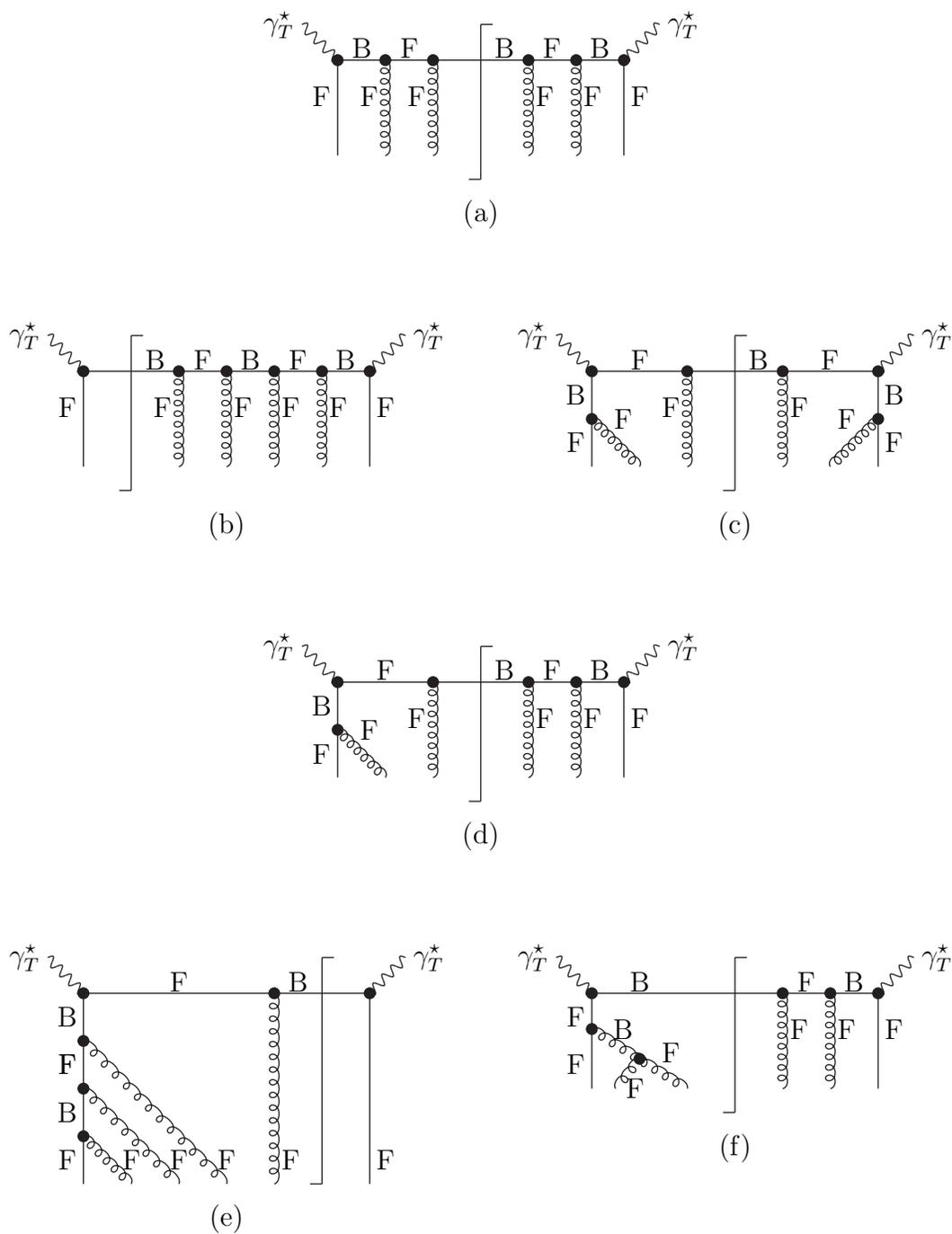
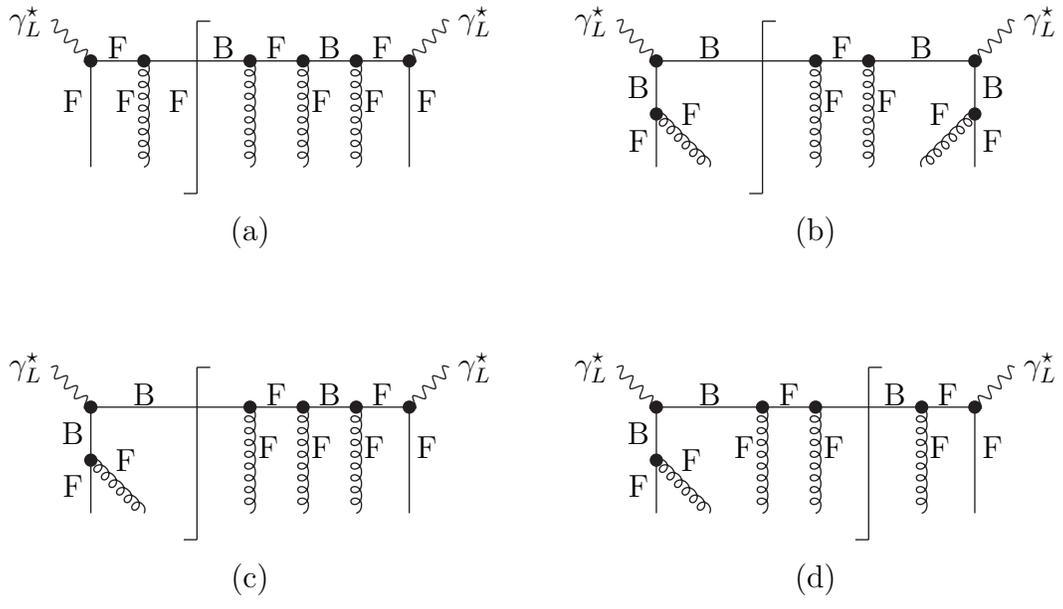
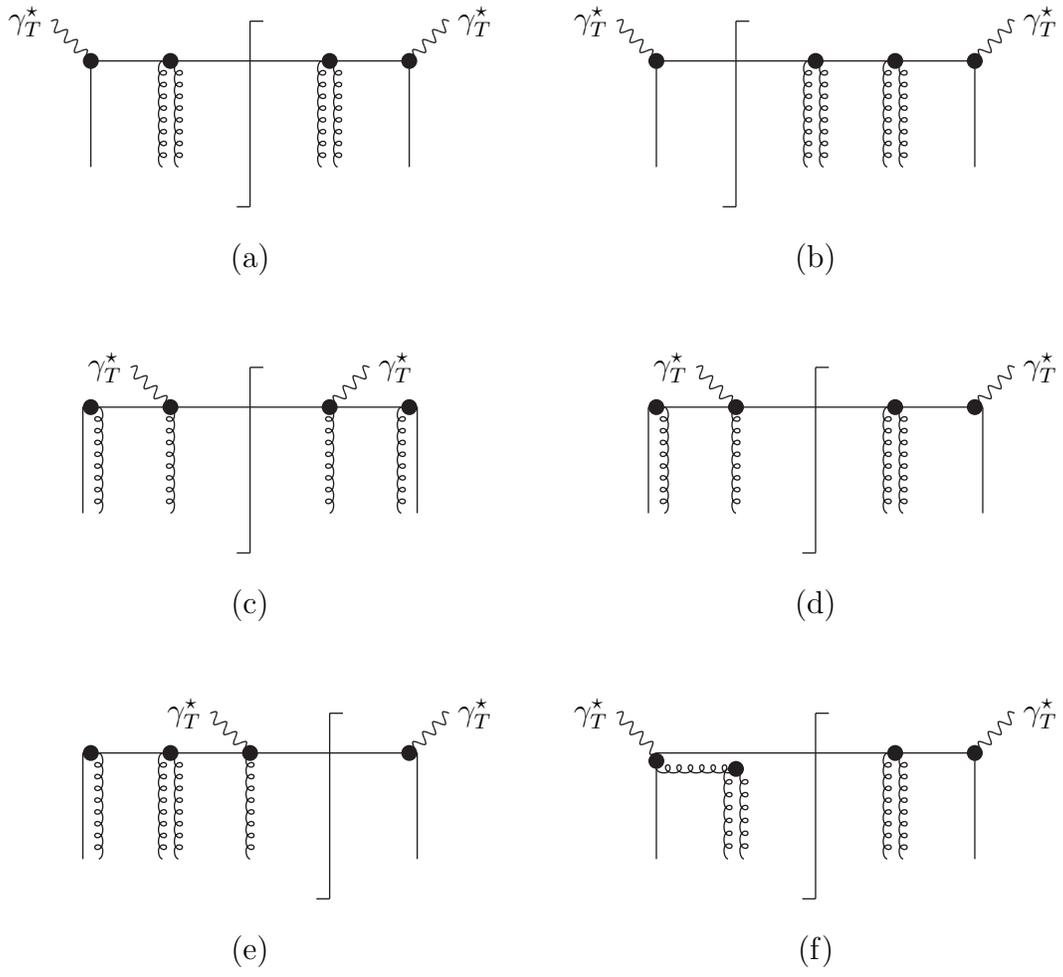
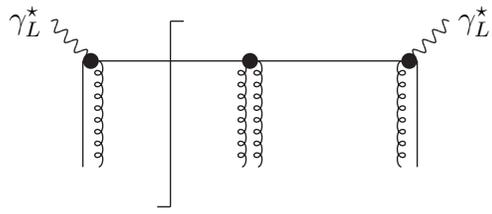
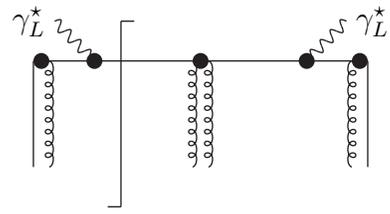
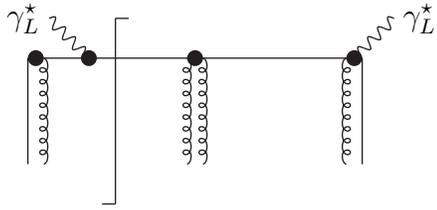
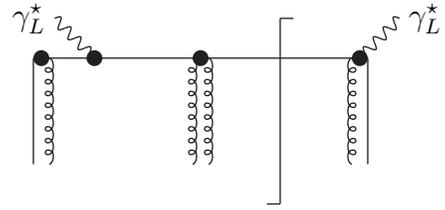
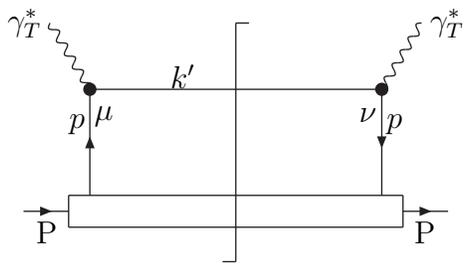
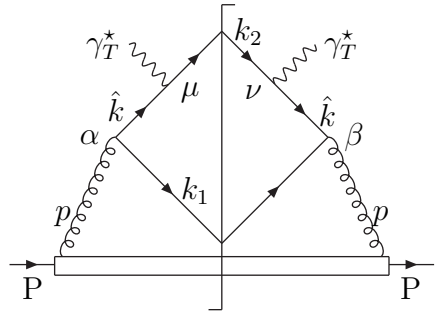
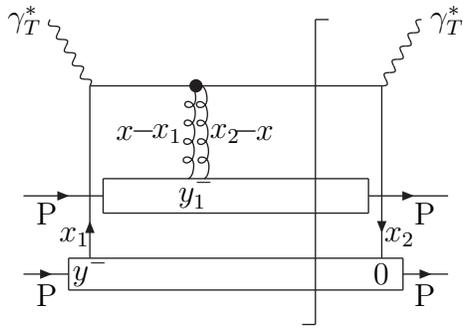
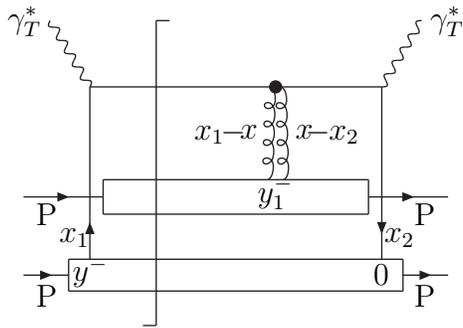
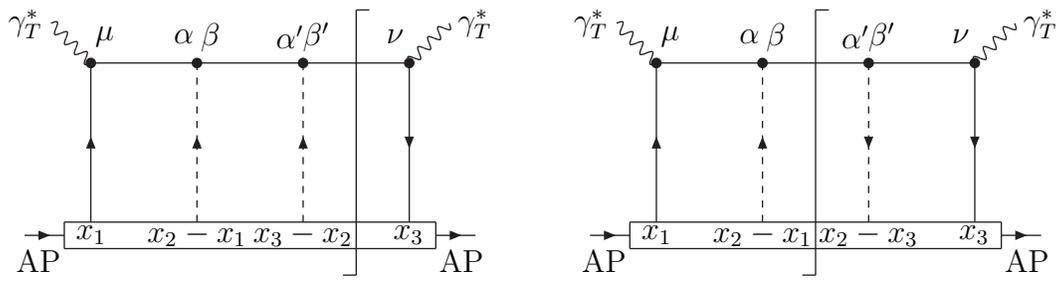
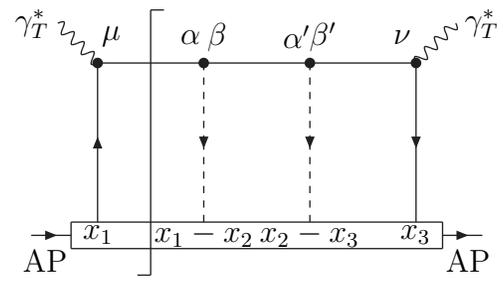
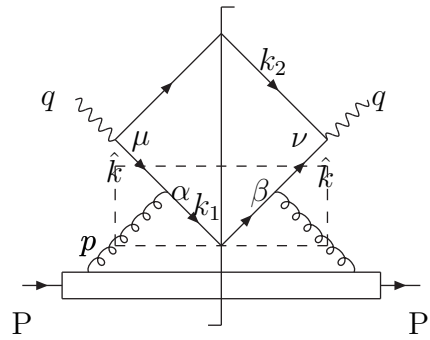
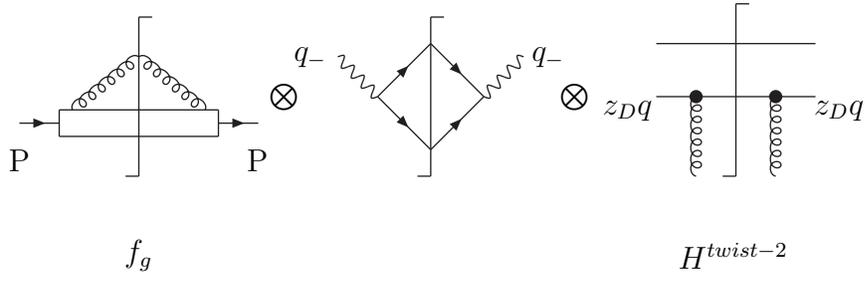
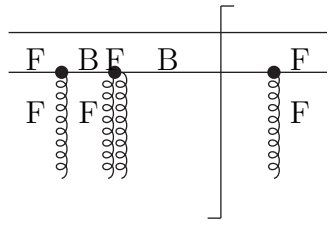
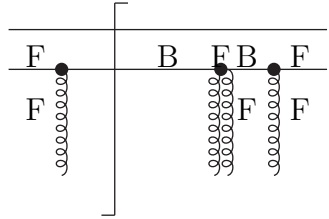
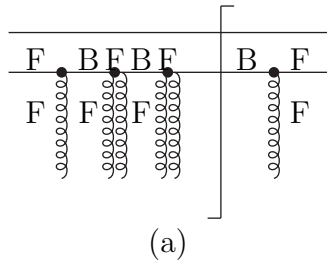
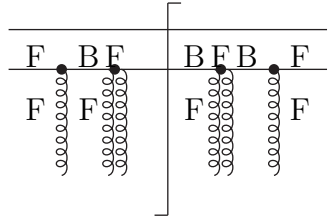
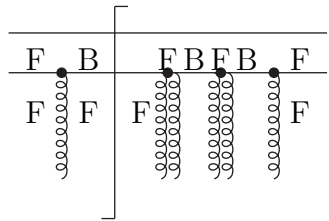
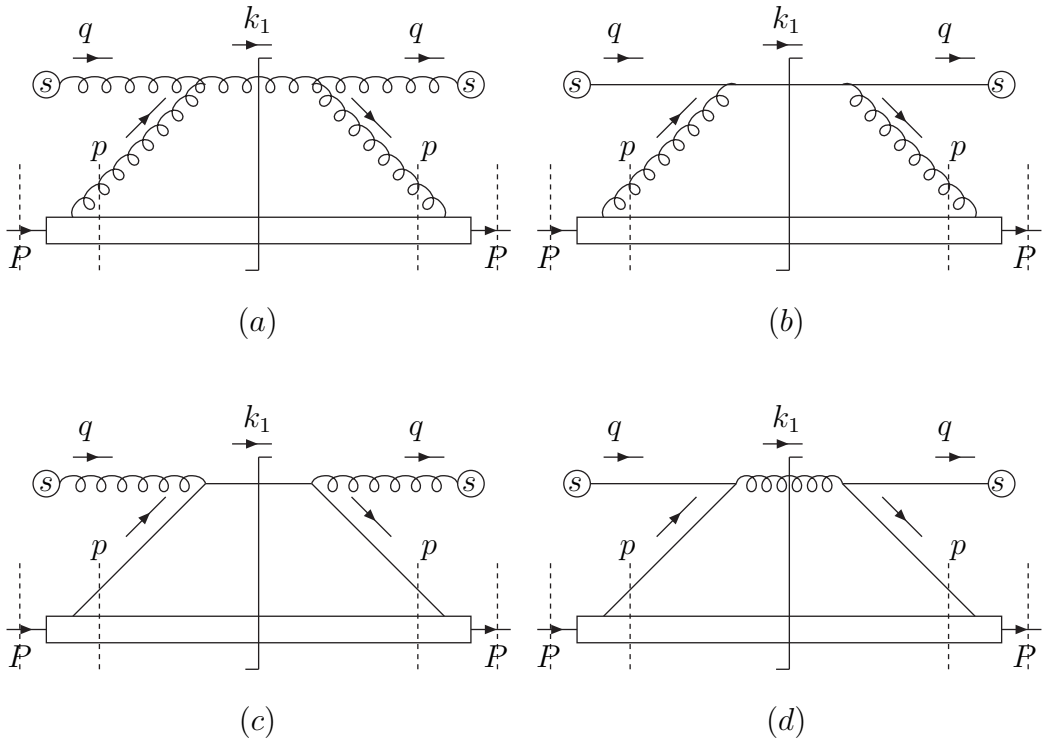
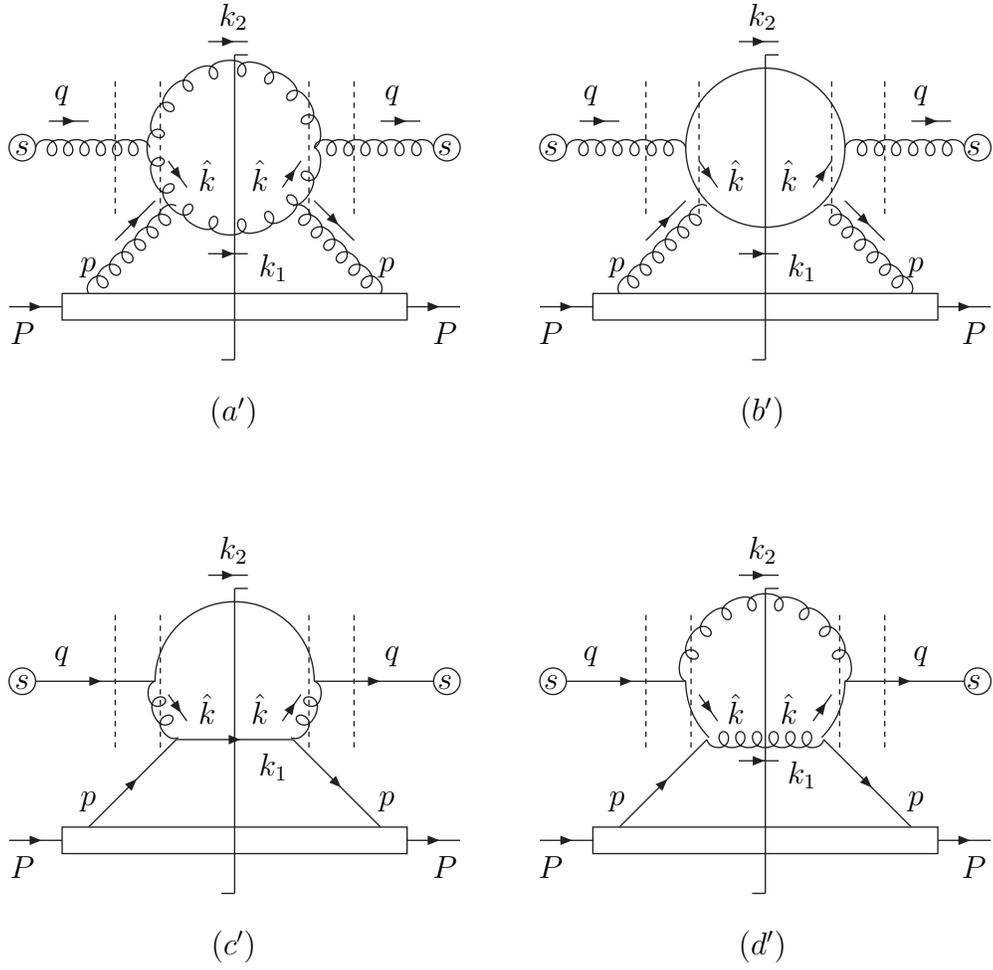
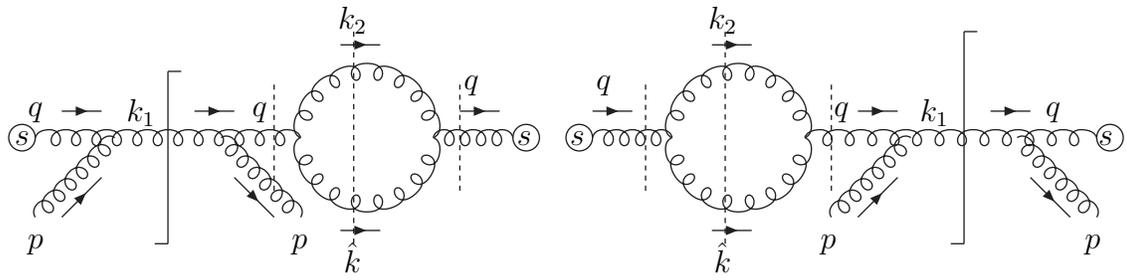
\end{document}